\documentclass[a4paper,11pt]{article}
\pdfoutput=1
\usepackage{jcappub}
\usepackage{mathtools}
\usepackage{ulem}
\usepackage{bm}
\usepackage{xcolor}

\renewcommand{\thesection}{\arabic{section}}

\def\laq{~\raise 0.4ex\hbox{$<$}\kern -0.8em\lower 0.62ex\hbox{$\sim$}~}
\def\gaq{~\raise 0.4ex\hbox{$>$}\kern -0.7em\lower 0.62ex\hbox{$\sim$}~}

\def\beq{\begin{equation}}
\def\eeq{\end{equation}}
\def\bea{\begin{eqnarray}}
\def\eea{\end{eqnarray}}
\def\bean{\begin{eqnarray*}}
\def\eean{\end{eqnarray*}}

\def\laq{~\raise 0.4ex\hbox{$<$}\kern -0.8em\lower 0.62ex\hbox{$\sim$}~}
\def\gaq{~\raise 0.4ex\hbox{$>$}\kern -0.7em\lower 0.62ex\hbox{$\sim$}~}

\def\be{\begin{equation}}
\def\ee{\end{equation}}

\def\beq{\begin{equation}}
\def\eeq{\end{equation}}
\def\bea{\begin{eqnarray}}
\def\eea{\end{eqnarray}}

\def \pa {\partial}

\newcommand{\Ocal}{\mathcal O}

\newcommand{\Hcal}{\mathcal H}
\newcommand{\Ical}{\mathcal I}
\newcommand{\Pcal}{\mathcal P}

\def\laq{~\raise 0.4ex\hbox{$<$}\kern -0.8em\lower 0.62ex\hbox{$\sim$}~}
\def\gaq{~\raise 0.4ex\hbox{$>$}\kern -0.7em\lower 0.62ex\hbox{$\sim$}~}

\def\beq{\begin{equation}}
\def\eeq{\end{equation}}
\def\bea{\begin{eqnarray}}
\def\eea{\end{eqnarray}}
\def\bean{\begin{eqnarray*}}
\def\eean{\end{eqnarray*}}

\def \pa {\partial}

\def \Ocal {\mathcal{O}}

\title{Relations between physical observables: what is better?}

\author{Giuseppe Fanizza}
\affiliation{Center for Theoretical Astrophysics and Cosmology,
Institute for Computational Science, University of Z\"urich, Winterthurerstrasse 190, CH-8057, Z\"urich, Switzerland}

\emailAdd{gfanizza@physik.uzh.ch}

\abstract{We investigate some possible relations between physical observables and estimate the ``cosmic variance'' which affects these measurements. We focus on redshift and angular-distance and we discuss the difference in considering the redshift as function of the angular-distance rather than the usually considered inverse relation. Already at linear level in metric perturbations, we find a significant difference. Indeed, even if both relations are led by source radial velocity for close enough sources, this effect is suppressed by 2 orders of magnitude in the redshift/angular-distance relation. This fact can significantly reduce the theoretical uncertainty for close sources already investigated in the literature for the angular-distance/redshift relation and open a new scenario for clarifying the tension in the measurement of $H_0$ from local sources rather than from the CMB.}

\begin{document}

\maketitle

\section{Introduction}
The relation between distances and redshift is one of most studied quantities in modern cosmology since the discovering of the acceleration of the Universe in the late 90's of the 20-th century \cite{Riess:1998cb,Perlmutter:1998np}. In particular the most widely investigated relation is the one between luminosity-distance and redshift. In fact the research in this regard concerns both exact inhomogeneous models (see \cite{Celerier:1999hp,Moffat:2005ii,Alnes:2005rw,Celerier:2009sv,Romano:2009xw,Fanizza:2014baa,Fleury:2014rea,Fleury:2016htl} and reference therein) and perturbative approaches.

The first attempt to describe linear perturbations in the luminosity-distance/redshift relation has been performed in 1987 \cite{Sasaki:1987ad}, when the acceleration of the Universe was still unrevealed. Once this accelerated expansion was discovered, the interest in describing inhomogeneous models became even larger. Indeed, the idea was to understand whether the inhomogeneities in the late Universe may mimic the accelerated expansion of the Universe via the so-called \textit{backreaction} on the background dynamics from the small scales averaging \cite{Buchert:2011sx}. Already at linear level some works tried to investigate the properties of the luminosity-distance/redshift relation \cite{Bonvin:2005ps,Pyne:2003bn} in both CDM and $\Lambda$CDM models. However, linear theory can be enough only when correlation functions and dispersions are studied. The real understanding of the impact of inhomogeneities on the averaged (or observed background) quantities requires at least non-linear perturbation theory to be adopted. The first attempt in this sense has been done in \cite{Barausse:2005nf}, where the perturbed luminosity-distance/redshift relation has been provided in the CDM model. This relation has been generalized to a generic dark energy model within General Relativity in a series of subsequent papers \cite{BenDayan:2012wi,Fanizza:2013doa} and then applied to the actual estimation of the averages over inhomogeneities on cosmological scales \cite{BenDayan:2012ct,BenDayan:2013gc,Ben-Dayan:2014swa}. Other non-linear evaluations of the luminosity-distance/redshift relation have been worked out independently in the literature \cite{Umeh:2012pn,Umeh:2014ana}\footnote{We redirect the reader interested in the comparison of these results to \cite{Marozzi:2014kua}, where the most general non-linear luminosity-distance/relation in presence of anisotropic stress is derived.}. Then some interesting applications in the understanding of the role of lensing in the CMB distance estimation have been analyzed \cite{Clarkson:2014pda,Bonvin:2015uha}. Finally the result of \cite{BenDayan:2012ct,BenDayan:2013gc} has been further investigated, in regard of the quantification of the bias in the Hubble diagram due to the adopted measure in the average process, in \cite{Fleury:2016fda}.

So far all the literature here revised considered photons that travel from the source to the observer position. However, whenever we discuss distances in an expanding Universe, we have to keep in mind that they really depend not only on the way in which we measure them, but also on the type of messenger involved in the specific measurement. Despite the fact that some proposals of new distance indicators with ultra-relativistic particles emitted from the source have been recently investigated from the theoretical point of view \cite{Zatsepin:1968kt,Fanizza:2015gdn,Stodolsky:1999kc,Fleury:2016mul}, most of the cosmological distance tools involve massless particles which travel from the source to the observer's laboratory. So far, the most widely adopted messengers are photons because they are stable particles which can almost freely travel along the Universe during its late time and due to the fact that they can be easily detected and studied in a satellite or ground-based experiment.

Photon traveling in the Universe is also very well understood from the theoretical point of view. Indeed it is well-known that their redshift $z$, angular-distance $d_A$ and luminosity-distance $d_L$ are related by the so-called Etherington reciprocity relation \cite{Etherington1933} as
\begin{equation}
d_L(z)=(1+z)^2d_A(z)\,.
\label{eq:Etherington}
\end{equation}
This relation is fully non-linear and holds as long as the photon number in a bundle of geodesics is conserved along its path. It basically states that $d_L$ and $d_A$ are proportional by a factor of squared redshift $z$ as long as both of them are expressed in terms of redshift itself. This means that both $d_A$ and $d_L$ in Eq. \eqref{eq:Etherington} refer to sources which lie on spheres at constant redshift along the past light cone of the observer. The general validity of this relation is a very useful tool and may provide a null-test about the validity of several proposed models for the late dynamics of the Universe \cite{DeBernardis:2006ii}. However, because of the inhomogeneities, constant redshift hyper-surfaces are not equal to the constant time-or-radius spheres. Moreover constant radius and constant time hyper-surfaces can be related between themselves only when the nature of the revealed messengers from the source to the observer is specified. Focusing on the background, for a matter of simplicity, we can unambiguously relate constant radius spheres with the constant time ones in a unique way only when we consider photons (or massless particles).

The difference among these three spheres is important and plays a role when we want to build relations between observables. Indeed, usually the observables are expressed in terms of the observed redshift. This approach implies that all the sources on a given sphere share the same redshift. Nevertheless we can express also the observables in terms of the observed angular-distance. In this case, all the sources will share the same $d_A$ on the given sphere. From the observational point of view, both procedures are legitimate.
\begin{figure}[ht!]
\centering
\includegraphics[scale=0.6]{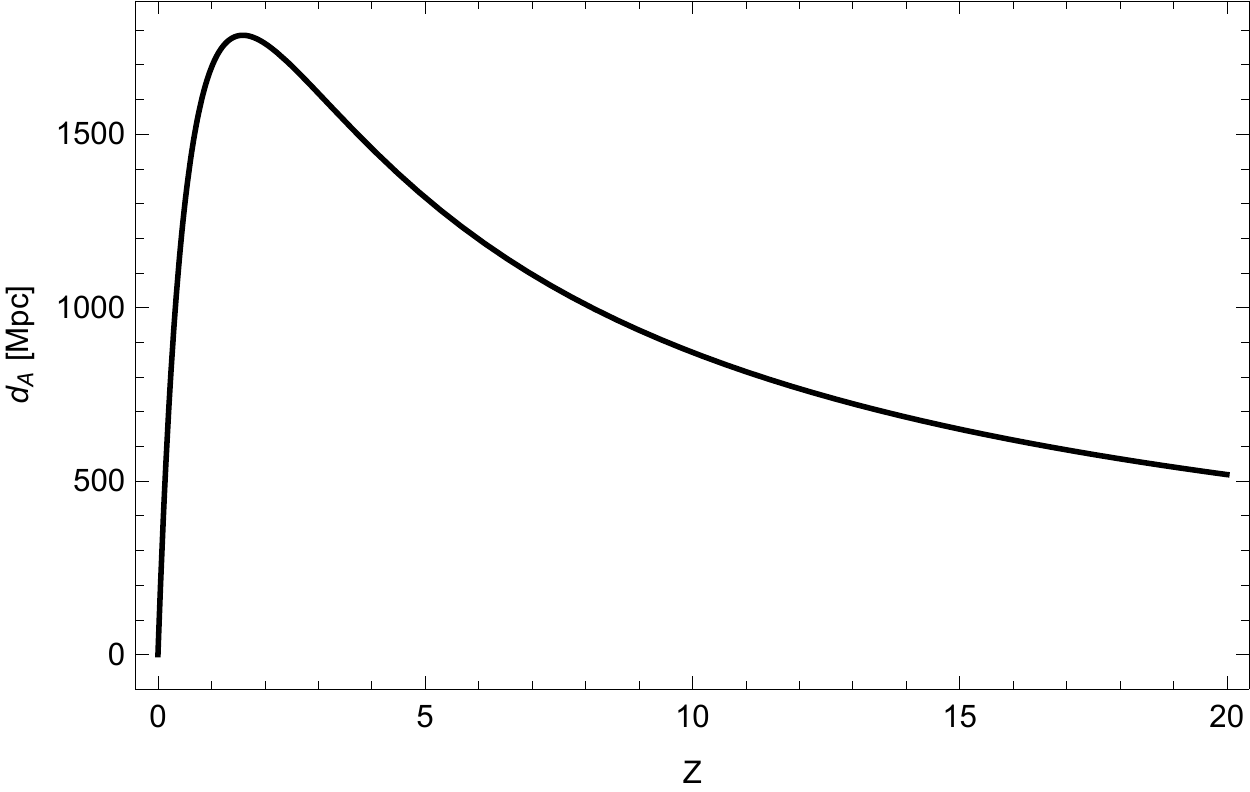}
\caption{Angular-distance/redshift relation for the background $\Lambda$CDM cosmology. As well-known, the relation between $d_A$ and $z$ is not invertible for any redshift. In particular, it exhibits a maximum around $z=1.6$ which corresponds to $d_A=1.8$ Gpc. For our purposes, throughout this paper we will only consider the monotonically increasing region $0\le z\le 1.5$.}
\label{fig:monotonically}
\end{figure}

However, from the theoretical point of view, redshift is usually taken as the independent variable. Among all the reasons for this choice there is the fact that the relation $d_A(z)$ on the background is defined for all redshifts (see Fig. \ref{fig:monotonically}). On the contrary, if we consider the inverse relation $z(d_A)$, we have to keep in mind that it can be studied only in ranges where the angular-distance/redshift relation is invertible. For instance, we can consider sources close enough ($z<1.5$), where the background relation is invertible and then study the difference in considering observables expressed in terms of equal-angular-distance sources rather than equal-redshift ones.

Because of this, the study that we are about to present in this paper is applicable only to late time measurements. However, even if constrained by this requirement, some very interesting features appear already in the chosen range. By looking at the explicit expressions in linear perturbation theory, for very close/young sources, the redshift/angular-distance relation exhibits a value for the dispersion which is significantly lower than the one for the usually adopted $d_A(z)$. This value is also competitive with the actual uncertainty of $H_0$ measurement from Planck data \cite{Ade:2013zuv}. Motivated by these reasons, in this paper we will quantify all these aspects and discuss the possible consequences in alleviating the tension between the local measurement of $H_0$ \cite{Riess:2011yx} and the value obtained from CMB observations \cite{Ade:2013zuv}.

This paper is then organized as follows. In Sect. \ref{sec:linear}, we discuss the difference in considering observables in terms of the constant angular-distance spheres rather than the equal redshift ones. For both cases we will evaluate the shift in time and position at the linear level and we will provide the linear expression for the perturbed relations $d_A(z)$ and $z(d_A)$. In Sect. \ref{sec:variance}, we will estimate the dispersion with respect to the background values for these relations due to the stochastic distribution of matter in the Universe between the sources and the observer and we will show the improvement that we gain when we consider $z(d_A)$ rather than $d_A(z)$. In Sect. \ref{sec:others} we will discuss some possible contaminations of the total dispersion from some sub-leading effects. Even if they can lead to some significant changes, their impact is not huge enough to alterate the behavior of the leading terms. Finally, in Sect. \ref{sec:summary} we will summarize and discuss the impact of our results for a better understanding of local measurements of $H_0$. Technical details about perturbative expressions, gauge invariance of the results and expansion in Fourier space are given in Apps. \ref{app:perturbed}, \ref{app:GI} and \ref{app:tec}.

\section{Relations between observables in linear perturbation theory}
\label{sec:linear}
For the purposes of this work, we just focus on linear perturbation theory and consider the line element as
\begin{equation}
ds^2=-a^2(\eta)\,\left(d\eta^2 -dr^2-r^2\,d\Omega^2 \right)+\delta g_{\mu\nu}dx^\mu dx^\nu\,,
\end{equation}
where $a$ is the scale factor as a function of the conformal time $\eta$, $r$ is a radial coordinate $d\Omega^2=d\theta^2+\sin^2\theta d\phi^2$ is the infinitesimal solid angle and $\delta g_{\mu\nu}$ are the metric perturbations with respect to the set of coordinates $x^\mu=\left( \eta,r,\theta,\phi \right)$. Moreover, we also expand
\begin{align}
d_L=&\bar d_L\left(1+\delta d_L\right)\nonumber\\
d_A=&\bar d_A\left(1+\delta d_A\right)\nonumber\\
1+z=&\left(1+\bar z\right)\left(1+\delta z\right)\,,
\label{eq:gaugedependent}
\end{align}
where $\bar d_A=r\,a(\eta)$, $1+\bar z=\frac{a(\eta_o)}{a(\eta)}$ and $\bar d_L=(1+\bar z)^2\bar d_A$ are the background relations and $\delta \Ocal$ is the perturbation related to the $\Ocal$ quantity. Because all the expressions in Eqs. \eqref{eq:gaugedependent} depend on the arbitrary set of coordinates $x^\mu$, they are also gauge-dependent. However, as shown in \cite{Fanizza:2018qux}, we can express a chosen observable $\Ocal$ in terms of the observed angles and another observable $\Pcal$ along the observer's past light-cone and get a gauge-invariant result. Hence, in the following, we firstly recall how to obtain the angular-distance in terms of the observed redshift and then we derive the inverse relation, i.e. $z(d_A)$, at linear level in perturbations.

\subsection{Angular-distance/redshift relation}
Here we briefly recall the procedure to relate angular-distance to redshift in linear perturbation theory\footnote{See \cite{Bonvin:2005ps,Scaccabarozzi:2017ncm} and also \cite{BenDayan:2012wi,Yoo:2017svj,Fanizza:2018qux} for extensions of this procedure to second and third \cite{Fanizza:2015swa} order in perturbation theory.}. Let us suppose that we want to express a given observable $\Ocal$ in terms of the observed redshift $z_{obs}$ of photons emitted by the source. Looking at the background expressions, we need to compute only the radial and time expansion around the emission position on the constant-redshift sphere, namely $\hat r_z$ and $\hat \eta_z$
\beq
\eta=\hat\eta_z+\delta\eta_z\qquad,\qquad r=\hat r_z+\delta r_z
\eeq
because background quantities do not depend on the angular coordinates. Then, linear perturbations are already expressed in terms of the observed angles. In this way, the conditions that we want to satisfy are two
\begin{align}
\left(1+\bar z\right)\left(1+\delta z\right)=\frac{a(\hat \eta_o)}{a(\hat\eta_z)}\left[1+\delta z-\Hcal_z\delta \eta_z+\Hcal_o\delta\eta_o\right]&\equiv \frac{a(\hat \eta_o)}{a(\hat\eta_z)}\equiv 1+z_{obs}\nonumber\\
\hat \eta_z+\hat r_z+\delta\eta_z+\delta r_z+\delta w&\equiv\hat\eta_z+\hat r_z\,,
\label{eq:dLz}
\end{align}
where $\delta\eta_o$ is the time lapse associated to the time as measured in the rest frame of the observer $\hat \eta_o$ \cite{Biern:2016kys} and $\delta w$ includes all the intrinsic perturbations of the past light-cone \cite{BenDayan:2012pp}. The first of Eqs. \eqref{eq:dLz} defines $\hat\eta_z$ as the time associated to the spheres at constant observed redshift and the second one relates the spheres at constant observed radial position of the source along the past light-cone to the observed redshift. In other words Eqs. \eqref{eq:dLz} define the deviations between the sphere at constant time/radius and the sphere at constant redshift. Indeed, even if topologically equivalent, all these spheres differ among themselves and the perturbations in these relations for time and radius are respectively given by solving Eqs. \eqref{eq:dLz}, i.e.
\begin{align}
\delta\eta_z=&\frac{1}{\Hcal_z}\left( \delta z+\Hcal_o\delta\eta_o \right)\nonumber\\
\delta r_z=&-\delta w-\frac{1}{\Hcal_z}\left( \delta z+\Hcal_o\delta\eta_o \right)\,,
\label{eq:shiftdAz}
\end{align}
where $\Hcal_z=\Hcal(\hat\eta_z)$ is the Hubble function evaluated at the redshift $z$. In this way, the linear angular-distance expressed in terms of the observed redshift of photons reads as
\begin{align}
d_A(z_{obs})=&\bar d_{A\,z}\left( 1+\delta d_A+\Hcal_z\delta\eta_z+\frac{\delta r_z}{\hat r_z} \right)\nonumber\\
=&\bar d_{A\,z}\left[ 1+\delta d_A+\left(1-\frac{1}{\hat r_z\,\Hcal_z}\right)\left(\delta z+\Hcal_o\delta\eta_o\right)-\frac{\delta w}{\hat r_z} \right]\nonumber\\
\equiv&\bar d_{A\,z}\left[ 1+\delta d_{A\,z} \right]\,,
\end{align}
where $\bar d_{A\,z}\equiv \hat r_z\,a(\hat\eta_z)=\left(\hat\eta_o-\hat\eta_z\right)a(\hat\eta_z)$ and all the perturbations are evaluated at $\eta=\hat\eta_z$ and $r=\hat r_z$. In this way we have defined $\delta d_{A\,z}$ as the \textit{linear perturbation of the angular-distance/redshift relation}. It is evident that $\bar d_A$ and $\bar d_{A\,z}$ are intrinsically different: indeed the former expresses the angular-distance in terms of constant time or radius surface whereas the latter entirely relates $d_A$ to the observed redshift \textit{along the observer's past light-cone}. Note also that $\delta d_{A\,z}$ is different from the simple perturbation of angular-distance. In fact it contains perturbations with respect to the sphere at constant observed redshift. Instead $\delta d_A$ just contains perturbations with respect to the two different spheres at constant time and radius. We can also understand this difference by looking at their gauge transformation properties: indeed $\delta d_A$ transforms as scalar under a gauge shift because $\eta$ and $r$ are just coordinates and then depends on the  adopted choice for them, whereas $\delta d_{A\,z}$ is gauge invariant (see App. \ref{app:GI}) and corresponds to a \textit{real measurable relation} between the observed angular-distance and the observed redshift, which is gauge independent.

Through the Etherington relation in Eq. \eqref{eq:Etherington}, we can also look at the relation between the observed luminosity-distance and the observed redshift. Indeed we have
\begin{align}
d_L(z_{obs})=&\left[\frac{a(\hat\eta_o)}{a(\hat\eta_z)}\left(1+\delta z-\Hcal_z\delta \eta_z+\Hcal_o\delta\eta_o\right)\right]^2\nonumber\\
&\times\bar d_{A\,z}\left[ 1+\delta d_A+\left(1-\frac{1}{\hat r_z\,\Hcal_z}\right)\left(\delta z+\Hcal_o\delta\eta_o\right)-\frac{\delta w}{\hat r_z} \right]\nonumber\\
=&\left( 1+z_{obs} \right)^2\bar d_{A\,z}\left[ 1+\delta d_A+\left(1-\frac{1}{\hat r_z\,\Hcal_z}\right)\left(\delta z+\Hcal_o\delta\eta_o\right)-\frac{\delta w}{\hat r_z} \right]\nonumber\\
\equiv&\bar d_{L\,z}\left( 1+\delta d_{L\,z} \right)\,,
\end{align}
where $\bar d_{L\,z}\equiv \left(1+z_{obs}\right)^2\,\bar d_{A\,z}$ and again all the quantities are evaluated at $\eta=\hat\eta_z$ and $r=\hat r_z$. Here too, we have defined the perturbation of the luminosity-distance/redshift relation as $\delta d_{L\,z}$. We then have that the perturbations of $d_L(z)$ and $d_A(z)$ are equal, i.e.
\begin{equation}
\delta d_{L\,z}=\delta d_{A\,z}\,.
\label{eq:2.9}
\end{equation}

In the following, we will discuss what happens when we evaluate the perturbation in the inverse relation, namely redshift/angular-distance relation.

\subsection{Redshift/angular-distance relation}
Just as done before for the constant redshift spheres, now we want to construct our observed shift with respect to the constant angular-distance sphere. To do this, now we expand
\beq
\eta=\hat\eta_d+\delta\eta_d\qquad,\qquad r=\hat r_d+\delta r_d\,,
\eeq
where $\hat\eta_d$ and $\hat r_d$ are time and radial position of the source measured on the constant $d_A$ spheres. Because of this, the shift between the constant $d_A$ sphere and the constant $\eta$ one, $\delta \eta_d$, and between the constant angular-distance hyper-surface and the constant $r$ one, $\delta r_d$, have to satisfy
\begin{align}
\bar d_A\left( 1+\delta d_A \right)=\hat r_d\,a(\hat \eta_d)\left( 1
+\delta d_A
+\frac{\delta r_d}{\hat r_d}
+\Hcal_d\,\delta\eta_d \right)
&\equiv\hat r_d\,a(\hat \eta_d)\equiv d_{A\,obs}\nonumber\\
\hat \eta_d+\hat r_d+\delta\eta_d+\delta r_d+\delta w&\equiv\hat\eta_d+\hat r_d\,.
\label{eq:dLdA}
\end{align}
Again, the second condition corresponds to imposing that we are detecting photons and their past light-cone coincides with the observed one. On the other hand, the first of Eqs. \eqref{eq:dLdA} defines the condition for our observed frame where the sources on a given sphere are taken at the same angular-distance. From these conditions, we get
\begin{align}
\delta\eta_d=&\frac{1}{1-\hat r_d\,\Hcal_d}\left(\hat r_d\,\delta d_A
-\delta w\right)\nonumber\\
\delta r_d=&-\frac{\hat r_d}{1-\hat r_d\,\Hcal_d}\,\delta d_A
+\left(\frac{1}{1-\hat r_d\,\Hcal_d}-1\right)\delta w\,,
\label{eq:shiftzDa}
\end{align}
where now $\Hcal_d\equiv\Hcal(\hat \eta_d)$ is the Hubble function given in terms of the time related to the observed angular-distance. The difference between the perturbation in Eqs. \eqref{eq:shiftdAz} and Eqs. \eqref{eq:shiftzDa} is due to the fact that the former refers to the perturbations with respect to the constant redshift sphere while the latter considers spheres at constant angular-distance. Even if both are constrained by the past light-cone of the observer, there is no reason why those shifts should agree already at linear order in perturbations of the metric.

Therefore we can expand all our observables in terms of the observed angular-distance. For the redshift we then get
\begin{equation}
z\left(d_{A\,obs}\right)
=\bar z_d\left( 1+\delta z+\Hcal_o\delta\eta_o-\Hcal_d\delta\eta_d \right)
\equiv\bar z_d\left( 1+\delta z_{d} \right)\,,
\end{equation}
where $1+\bar z_d\equiv \frac{a(\hat \eta_o)}{a(\hat \eta_d)}$. Here $\delta z_d$ is the \textit{perturbation of the redshift/angular-distance relation}, which is different from the perturbation of redshift $\delta z$. Moreover, we underline that $\delta z_d$ is gauge invariant (see again App. \ref{app:GI}) because it is the perturbation of a relation between physical observables. The reason is the same that we pointed out for $\delta d_{A\,z}$. In the following, we will discuss some interesting consequences of these results.

\section{Comparison between the results}
\label{sec:variance}
In this section, we discuss the consequences of which kind of relation between observables is considered. First of all, let us write the explicit form in terms of the standard perturbation theory of the two terms of our interest. We have
\begin{equation}
\delta d_{A\,z}=\delta d_A-\frac{\delta w}{\hat r_z}
+\frac{\hat r_z\,\Hcal_z-1}{\hat r_z\,\Hcal_z}\,\left(\delta z+\Hcal_o\,\delta\eta_o\right)
\label{eq:DLZ}
\end{equation}
and
\begin{equation}
\delta z_d=\frac{\hat r_d\,\Hcal_d}{\hat r_d\,\Hcal_d-1}\left(\delta d_A
-\frac{\delta w}{\hat r_d}\right)
+\delta z+\Hcal_o\delta\eta_o\,.
\label{eq:DLDA}
\end{equation}
We notice that in both $\delta d_{A\,z}$ and $\delta z_d$ we have the combinations $\delta z+\Hcal_o\delta\eta_o$ and $\delta d_A-\delta w/\hat r$ and their coefficient are such that in both cases the gauge invariance of the result is obtained, as explicitly shown in App. \ref{app:GI}. The complete expressions for $\delta d_{A\,z}$ and $\delta z_d$ in the longitudinal gauge are reported in App. \ref{app:perturbed}. All the numerical results in the following will refer to this gauge choice. Working in a given gauge is perfectly allowed once the gauge invariance of the result is proven.

Hence, first of all, we recall that both expressions are dominated by lensing convergence $\kappa$, which appears in $\delta d_A$, and, for closer/younger sources, by their radial velocity $v_{\rVert}$, which is in the $\delta z$. However, there is an important difference between these behaviors. Indeed, the factors which multiply those terms are different. This can lead to important differences in the estimators for one relation rather than the other. Indeed, in Eq. \eqref{eq:DLZ} we notice that $\delta d_A$ is multiplied by a constant factor, whereas $\delta z$ is modulated by a factor which suppresses its contribution when $\hat r_z\sim\Hcal_z^{-1}$, namely when the distance of the source is comparable with the horizon. On the other hand, this factor is enlarged when $\hat r_z \ll \Hcal_z^{-1}$. Because of this, the angular-distance/redshift relation is dominated by lensing at higher redshift and by Doppler effect for younger sources \cite{Bonvin:2005ps}. Some possible consequences of this fact have been already investigated in the literature \cite{BenDayan:2013gc,Ben-Dayan:2014swa}, in particular in the estimation of some theoretical bias in the local value $H_0$ local.

On the other hand, Eq. \eqref{eq:DLDA} exhibits quite different properties. Indeed, in the $z(d_A)$ relation, the contribution from source radial velocity is no longer magnified for closer sources. Most likely, it is equally weighted, regardless of the distance between the source and the observer. Lensing convergence, instead, is suppressed when $\hat r_d \ll \Hcal_d^{-1}$ and it gets more important as long as the source is closer to the horizon scale $\hat r_d \sim \Hcal_d^{-1}$. In particular, the first of these regimes is important for our discussion. On top of this direct measurements of angular-distance are really challenging for faint sources because this requires that we must have access to a statistically significant number of objects which are huge enough to make reasonable a measurement of their size\footnote{Here we stress that we refer to direct measurement of the angular-distance. Indeed, an indirect measurement of $d_A$ could be performed through the luminosity-distance, thanks to the Etherington relation. However, this holds only when everything is expressed in terms of the constant redshift sphere, which is not the case for $z(d_A)$.}. Moreover, in order to consider the relation $z(d_A)$ at linear-order, we have to be sure that the relation $d_A(z)$ is invertible already at the background level. Because $d_A(z)$ is monotonically increasing only for $z<1.5$ (as shown in Fig. \ref{fig:monotonically}), we consider only this regime in the rest of our paper. In this range, $\hat r_z\sim\Hcal_z^{-1}$ just when $z\sim 1.5$.

Hence, in an inhomogeneous Universe we have that each line-of-sight can be view as a particular realization of the perturbations. We can then integrate the perturbations along each line-of-sight for several sources and invoke the \textit{ergodic theorem}. In this way, the average over all the possible directions is meant to be equivalent to the average over several realization of inhomogeneities. Because of this, inhomogeneities will induce a variance $\sigma^2$ on the simple background relations. In the following, we will indicate the \textit{average over directions} with $\langle\dots\rangle$ and the \textit{ensamble average} over different realizations of the perturbations with $\overline{\textcolor{white}{l}\dots\textcolor{white}{l}}$ (definitions and technical details about them are reported in App. \ref{app:tec}). At linear order these averages are commutative operations and do not involve any perturbations in the measure. The non-commutative behavior between these average procedures is a pure second-order effect \cite{BenDayan:2012pp,Bonvin:2015kea,Fleury:2016fda} and here can be safely neglected. Moreover, the evaluation of second-order quantities, as backreaction, requires the estimation of perturbations also in the measure adopted for the average \cite{BenDayan:2012pp}. Let us notice that both rigorous \cite{Gasperini:2011us} and phenomenological \cite{Fleury:2016fda} proposals for the averages may involve the $d^2_A$ in the measure of the average itself. It is then interesting to notice that for $z(d_A)$ this part of the measure should not be affected by the perturbations, because the average is performed over the sphere at constant angular-distance. This is not the case for the $d_A(z)$ relation, where the integral is performed over the sphere at constant redshift. The consequence of this difference will be investigated in a subsequent work related to second-order estimators.

Hence, focusing on linear-order, for the luminosity-distance/redshift relation we then have
\begin{equation}
\sigma^2_{d_{A\,z}}\equiv\overline{\langle \delta d^2_{A\,z} \rangle}=
\overline{\langle \delta d^2_A \rangle}
+\frac{1}{\hat r_z^2}\overline{\langle \delta w^2 \rangle}
+\left(\frac{\hat r_z\,\Hcal_z-1}{\hat r_z\,\Hcal_z}\right)^2\,\overline{\langle\delta z^2\rangle}+\text{cross terms}\,
\end{equation}
while, for the redshift/angular-distance relation the result is
\beq
\sigma^2_{z_d}\equiv\overline{\langle \delta z^2_d \rangle}=
\left(\frac{\hat r_d\,\Hcal_d}{\hat r_d\,\Hcal_d-1}\right)^2\overline{\langle \delta d^2_A \rangle}
+\left(\frac{\Hcal_d}{\hat r_d\,\Hcal_d-1}\right)^2 \overline{\langle \delta w^2 \rangle}
+\overline{\langle\delta z^2\rangle}+\text{cross terms}\,.
\eeq
From the explicit form of their perturbative expressions, $\delta d_A$ is dominated by lensing convergence $\kappa$ while $\delta z$ is mostly affected by radial Doppler velocity of the source $v_{\rVert}$ so both variances can be written as
\begin{align}
\sigma^2_{d_{A\,z}}=&\overline{\langle \kappa^2 \rangle}
+\left(\frac{\hat r_z\,\Hcal_z-1}{\hat r_z\,\Hcal_z}\right)^2\overline{\langle v^2_{\rVert} \rangle}
+\text{others}
\equiv\left(\sigma^2_{d_{A\,z}}\right)_\kappa+\left(\sigma^2_{d_{A\,z}}\right)_{v_{\rVert}}+\text{others}
\nonumber\\
\sigma^2_{z_d}=&\left(\frac{\hat r_d\,\Hcal_d}{\hat r_d\,\Hcal_d-1}\right)^2\,\overline{\langle \kappa^2 \rangle}
+\overline{\langle v^2_{\rVert} \rangle}
+\text{others}
\equiv\left(\sigma^2_{z_d}\right)_\kappa+\left(\sigma^2_{z_d}\right)_{v_{\rVert}}+\text{others}\,.
\label{eq:variances}
\end{align}
In these expressions, we are neglecting the contribution from the cross-term $\overline{\langle \kappa v_\rVert \rangle}$. Even if this terms involves three spatial derivatives of the gravitational potential, its amplitude is 2-3 orders of magnitude lower than the auto-correlation of $\kappa$ and $v_\rVert$ so it can be safely neglected.

\begin{figure}[ht!]
\centering
\includegraphics[scale=0.59]{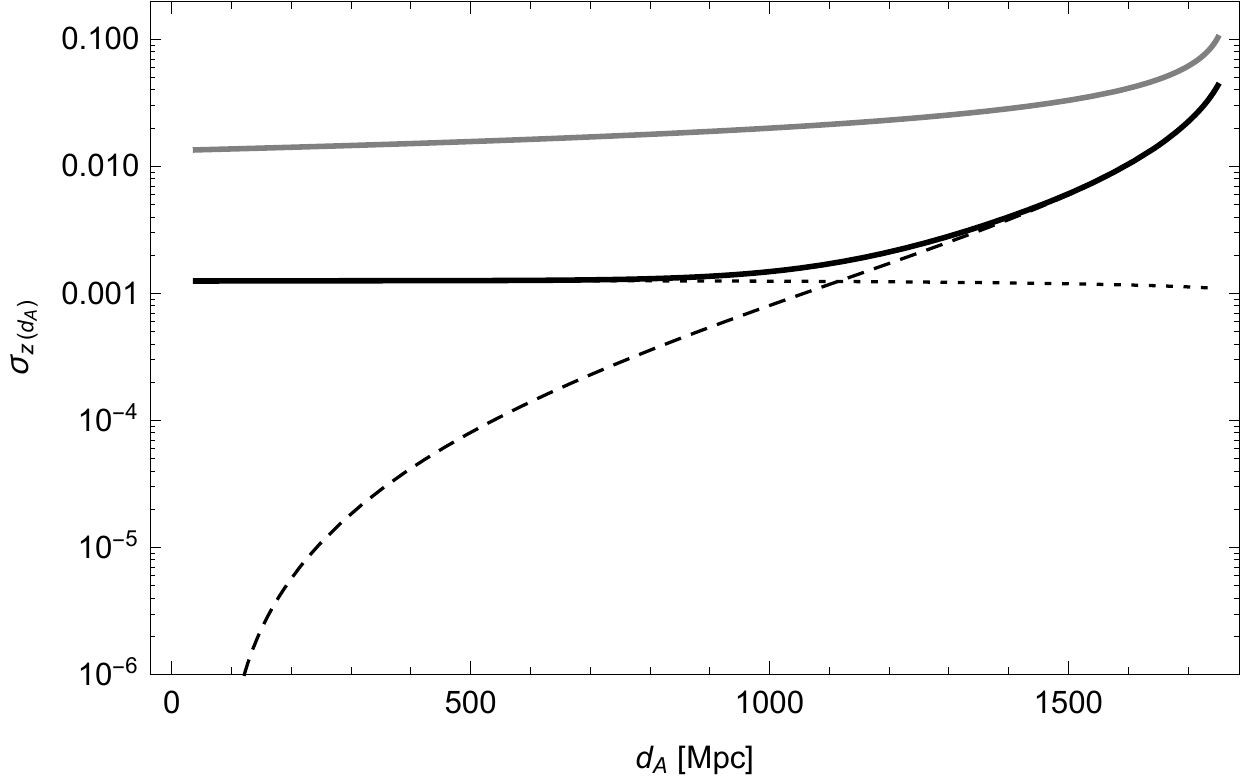}
\includegraphics[scale=0.6]{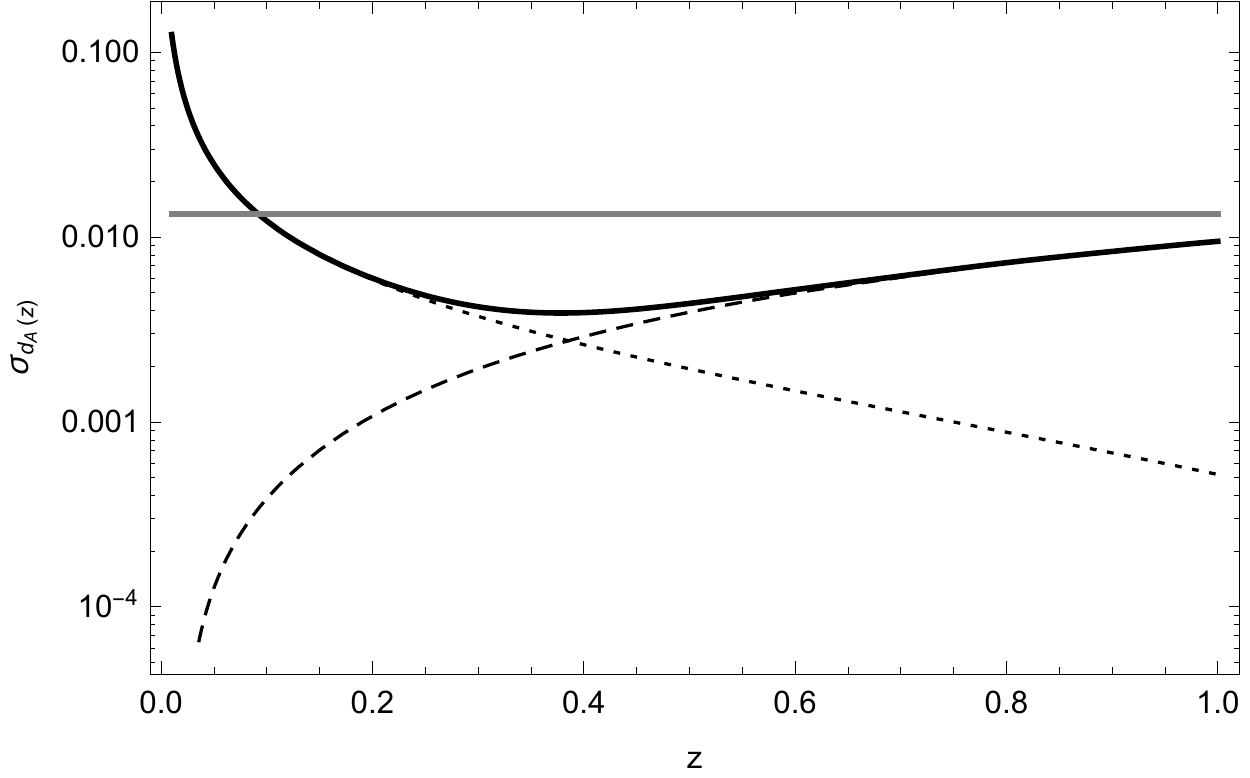}
\caption{Dispersion for the redshift/angular-distance relation (left panel) and for the angular-distance/redshift relation (right panel) due to the lensing convergence $\kappa$ (dashed lines) and source radial velocity $v_{\rVert}$ (dotted lines). Solid black lines represent the quadratic sum of these effects, i.e. $\sigma=\sqrt{\sigma_\kappa^2+\sigma^2_{v_{\rVert}}}$, while solid gray lines consider the dispersion due to the error in the measured $H_0$ from Planck data. Those plots exhibit the same qualitative behaviors: indeed their are both dominated by lensing for fainter/older sources, whereas radial velocity of the sources themselves is dominant in closer/later regimes. However, for the relation $z(d_A)$ the dispersion for closer sources is lower than its estimation for $d_A(z)$ by almost 2 orders of magnitude.}
\label{fig:plots}
\end{figure}

Already at a first sight, we notice that these variances are quantitatively different, even if sourced by the same relativistic effects. In details, in Fig. \ref{fig:plots} we show the total behavior of the dispersions $\sigma\equiv\sqrt{\sigma^2}$ (solid black lines) and we outline the two different contributions from lensing convergence (dashed curves) and radial velocity (dotted lines).
The first consideration that we immediately see is that in both panels the same qualitatively behavior is shown: in particular, radial Doppler velocity is relevant for closer/younger sources. Instead, dispersions for older/fainter sources are dominated by lensing. This result was already known for the luminosity-distance/redshift relation. Indeed we have a qualitative agreement with the results presented in \cite{BenDayan:2013gc,Biern:2016kys}. This agreement is very good also quantitatively for small redshifts whereas it gets less accurate for higher redshift. The reason can be related to the fact that here we limit our analysis to the linear power spectrum for the gravitational potential \cite{Eisenstein:1997ik} (see Appendix \ref{app:tec} for numerical details). This implies that we can only treat scales up to $k_{UV}=1\,h$ Mpc$^{-1}$. Instead the results presented in \cite{BenDayan:2013gc} consider the non-linear power spectrum from HaloFit \cite{Smith:2002dz,Takahashi:2012em}. This allows them to extend their integration domain in Fourier space up to $k_{UV}=30\,h$ Mpc$^{-1}$, where non-linear evolution of modes becomes important. However, both results catch the same order of magnitude of the effect. Anyway the difference is important only at higher redshifts. Hence the analysis made here remains valid.

The fact that the same hierarchy between lensing convergence and source velocity holds also for the redshift/angular-distance relation is because lensing is usually the leading effect when compared to Doppler. Indeed the latter involves only one spatial derivative of the Bardeen potential whereas convergence counts two angular derivatives of the gravitational potential. The dominance of lensing may happen just in some given regime close enough to the observer such that the integrated effects along the line-of-sight (lensing is one of them) are negligible with respect to local effects as the local motion of the source.

What is more interesting is the following thing: in the regime where $v_{\rVert}$ is dominant, the dispersion of $z(d_A)$ is almost 2 orders of magnitude lower than the one for $d_A(z)$. This numerical difference is explained by Eqs. \eqref{eq:variances}. Indeed, if we consider $\epsilon\equiv\hat r_z\,\Hcal_z$ small enough, we have that the factor in front of $\overline{\langle v^2_{\rVert} \rangle}$ for the dispersion of the angular-distance/redshift relation is $1-\left(\hat r_z\,\Hcal_z\right)^{-1}=1-\epsilon^{-1}\approx -\epsilon^{-1}$. In this way for close/young sources, i.e. when $\kappa$ is negligible, $\sigma^2_{d_{A\,z}}\approx \epsilon^{-2}\,\overline{\langle v^2_{\rVert} \rangle}$ whereas $\sigma^2_{z_d}\approx \overline{\langle v^2_{\rVert} \rangle}$ and this leads to $\sigma_{z_d}\sim\epsilon\,\sigma_{d_{A,z}}$. Hence, because nowadays we can take $\Hcal^{-1}_z=\Hcal^{-1}_o\sim 10^3$ Mpc, if we consider distances of order $\hat r_z\sim10$ Mpc\footnote{Note that these distances roughly correspond to a range in redshift $z=0.01,0.05$.} then $\epsilon\sim 10^{-2}$ which is accordance with what numerically shown.  This indicates that a measurement for $z(d_A)$ from local sources is significantly less affected by theoretical uncertainty.

This difference in order of magnitude can have consequences for the possible better estimation of some cosmological parameters. In particular, we focus on the estimation of $H_0$. From Planck measurements, we have that the actual measured value is $H_0\pm\Delta H_0=67.8\pm 0.9$ Km/s Mpc$^{-1}$. Let us then define the variance associated to this experimental uncertainty as
\beq
\left(\sigma^{H_0}_{\Ocal_\Pcal}\right)^2=\left(\frac{\Ocal_\Pcal^{H_0=67.8+0.9}-\Ocal_\Pcal^{H_0=67.8-0.9}}{2\,\Ocal_\Pcal^{H_0=67.8}}\right)^2
\eeq
where $\Ocal$ and $\Pcal$ can be either $d_A$ or $z$ and $\Ocal_\Pcal$ is their \textit{background} relation. An analytical attempt of this estimation can be done exactly for the angular-distance/redshift relation. Indeed we can write without any lack of generality
\beq
\bar d_A(\bar z)=\frac{1}{H_0}f(\bar z)\,.
\eeq
In this way, the change in the relation due to a shift of $H_0$ will be simply given by
\beq
d_{A\,z}^{\,H_0\pm\Delta H_0}=\bar d_A(\bar z)\pm\Delta H_0\,\pa_{H_0}\bar d_A(\bar z)=\bar d_A(\bar z)\mp\frac{\Delta H_0}{H^2_0}f(\bar z)\,,
\eeq
which leads to
\beq
\left(\sigma^{H_0}_{d_{A\,z}}\right)^2=\left(\frac{\Delta H_0}{H_0}\right)^2=\left(1.3\times10^{-2}\right)^2\,.
\eeq
This variance then remains constant and is in perfect agreement with the numerical results shown in Fig. \ref{fig:plots} (solid gray lines). On the other hand, $\bar z(\bar d_A)$ can be written as
\beq
\bar z\left( \bar d_A \right)=f^{-1}\left(H_0\,\bar d_A\right)\equiv g\left( H_0\,\bar d_A\right)\,,
\eeq
so its shift due to a change in $H_0$ reads as
\beq
z_d^{\,H_0\pm\Delta H_0}=\bar z(\bar d_A)\pm\Delta H_0\,\pa_{H_0}\bar z(\bar d_A)=\bar z(\bar d_A)\pm\frac{\Delta H_0}{H_0}\,g'\left( H_0\,\bar d_A \right)\,H_0\,\bar d_A\,,
\eeq
where $g'(x)\equiv dg(x)/dx$, which leads to
\beq
\left( \sigma^{H_0}_{z_d} \right)^2=\left(\frac{\Delta H_0}{H_0}\,\frac{g'\left( H_0\,\bar d_A \right)}{g\left( H_0\,\bar d_A \right)}\,H_0\,\bar d_A\right)^2
=\left(1.3\times10^{-2}\right)^2\,\left(\frac{g'\left( H_0\,\bar d_A \right)}{g\left( H_0\,\bar d_A \right)}\,H_0\,\bar d_A\right)^2\,.
\eeq
This variance is more involved than the one for the inverse relation. A first obviuos difference that we notice is that it is distance-dependent. We then investigate its behavior for near sources such that $H_0\,\bar d_A\ll 1$. Because $H_0^{-1}\sim 10^3$ Mpc, this regime is valid until $\bar d_A \sim 10^2$ Mpc. For this kind of sources, $\bar z\sim 10^{-2}$, so $f(x)\approx x$ and then $g(x)=f^{-1}(x)=x^{-1}$. These analytical considerations allow us to say that $x\,g'(x)/g(x)=1$ and then $\left( \sigma^{H_0}_{z_d} \right)^2\approx\left(\sigma^{H_0}_{d_{A\,z}}\right)^2=\left(1.3\times10^{-2}\right)^2$, again in good agreement with the numerical results shown in Fig. \ref{fig:plots}.

The huge cosmic variance associated to the dispersion of sources velocity for local measurement has been addressed as one possible way to understand the tension between local measurement of $H_0$ and the one from CMB. In particular, it has been estimated \cite{Ben-Dayan:2014swa} that this large dispersion leads to some theoretical systematics in the evaluation of $H_0$ from local supernovae which is in the range $(1.6,2.4)$ Km/s Mpc$^{-1}$. On the contrary, the fact that this variance is 2 orders of magnitude lower for measurements of the redshift/angular-distance relation indicates that $z(d_A)$ is much less affected by this cosmic variance. Therefore a direct measurement of this inverse relation may exhibit a significantly different behavior in the estimation of $H_0$ from local sources because the direct comparison between this proposed estimation of $H_0$ and the one from CMB analysis will be free from the theoretical bias due to the velocity of local sources. If this were the case, this would also help in understanding whether the tension between $H_0$ local and $H_0$ CMB is truly due to the cosmic variance, just as explained in \cite{Ben-Dayan:2014swa}.

\section{Contributions from other effects}
\label{sec:others}
So far, we have only considered the effects due to lensing convergence and source Doppler velocity. This is justified by the fact that they have the highest number of spatial derivatives. In this section, we will discuss whether other effects may play a crucial role in other important regimes.
\subsection{Observer velocity}
The first effect that we want to consider is due to the peculiar velocity of the observer as generated by the cosmological perturbations along its world-line. We do not refer to the real velocity due to our motion within the non-linear gravitational potential of local structures as the Milky Way or the Solar System because it has no cosmological origin. Because of this, the proper estimation of this effect should be carried out with some exact estimations out of the cosmological perturbation theory framework. Having this in mind, the effect that here we call $v_{\rVert\,o}\equiv v_{\rVert}(\eta_s=\eta_o)$ is just due to the motion of a geodesic observer along a given set of perturbations \textit{on cosmological scales}. How to observationally disentangle those two different motions is a non-trivial task and it is well beyond the purpose of this work.

Let us then consider the peculiar velocity of the observer $v_{\rVert\,o}$. Differently from $v_{\rVert}$, it appears in both $\delta d_A$ and $\delta z$ (see Eqs. \eqref{eq:da} and \eqref{eq:z}), leading to the following expressions
\beq
\left(\delta d_{A\,z}\right)_{v_{\rVert\,o}}=\frac{1}{\hat r_z\Hcal_z}\,v_{\rVert\,o}\qquad\,\qquad
\left(\delta z_d\right)_{v_{\rVert\,o}}=\frac{1}{\hat r_d\Hcal_d-1}\,v_{\rVert\,o}\,.
\eeq
While the angular-distance/redshift relation is dramatically affected by this term as $\hat r_z\ll \Hcal_z^{-1}$, we notice that it remains constant in the same regime for $\delta z_d$. This is already an interesting behavior to be pointed out. On the other hand, $\delta z_d$ gets significantly affected by this term when $\hat r\sim\Hcal^{-1}$.
\begin{figure}[ht!]
\centering
\includegraphics[scale=0.6]{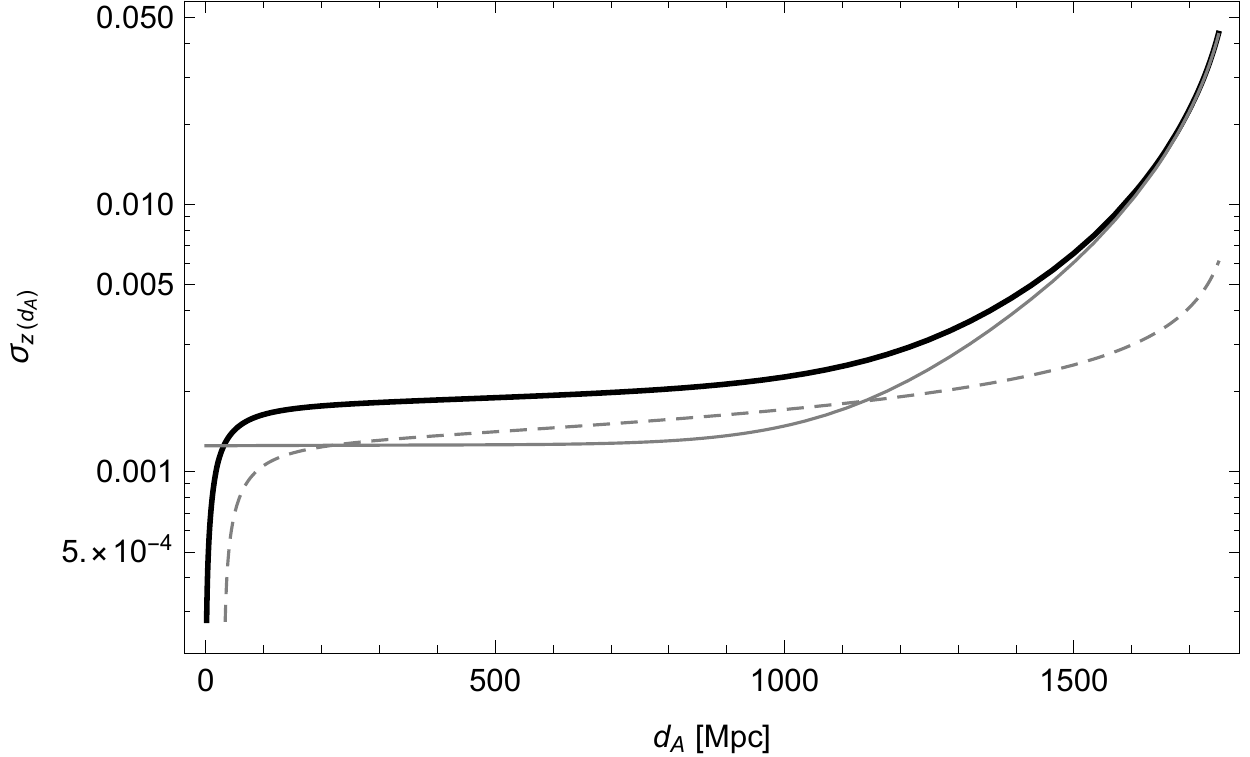}
\includegraphics[scale=0.605]{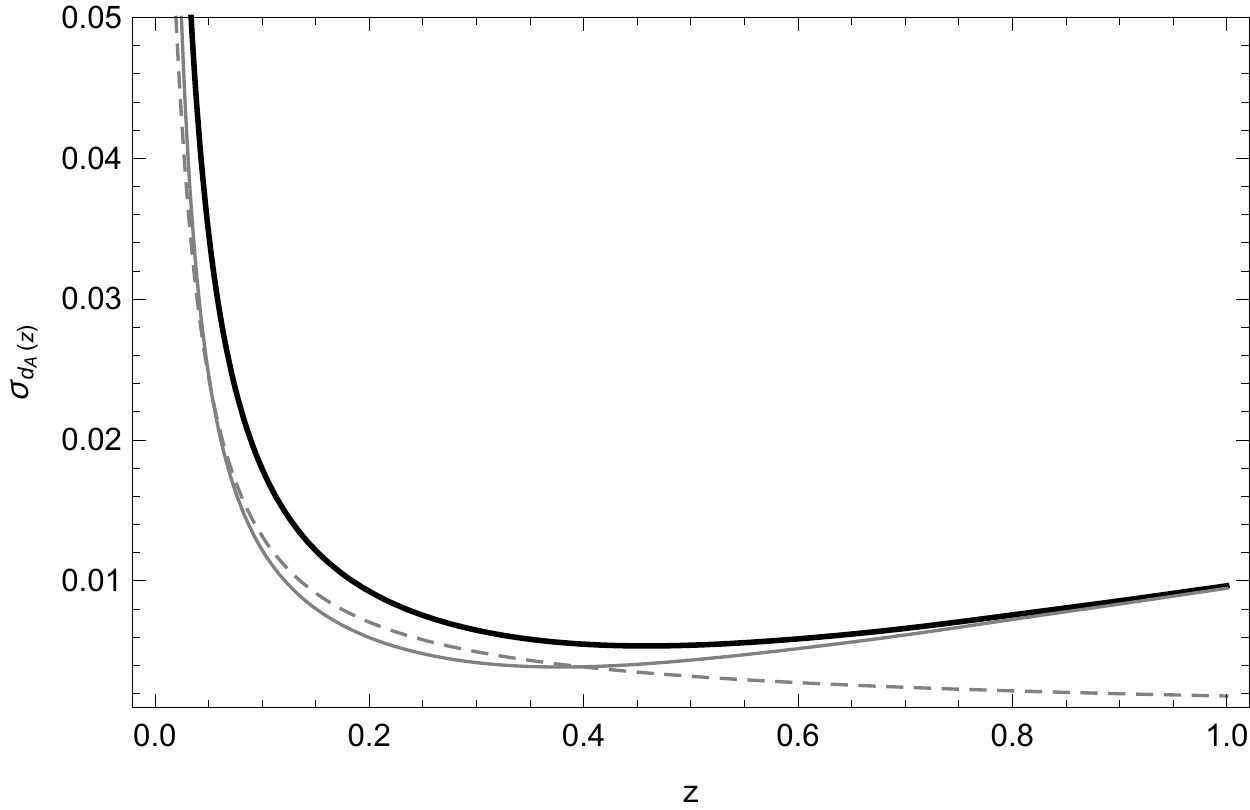}\\
\includegraphics[scale=0.6]{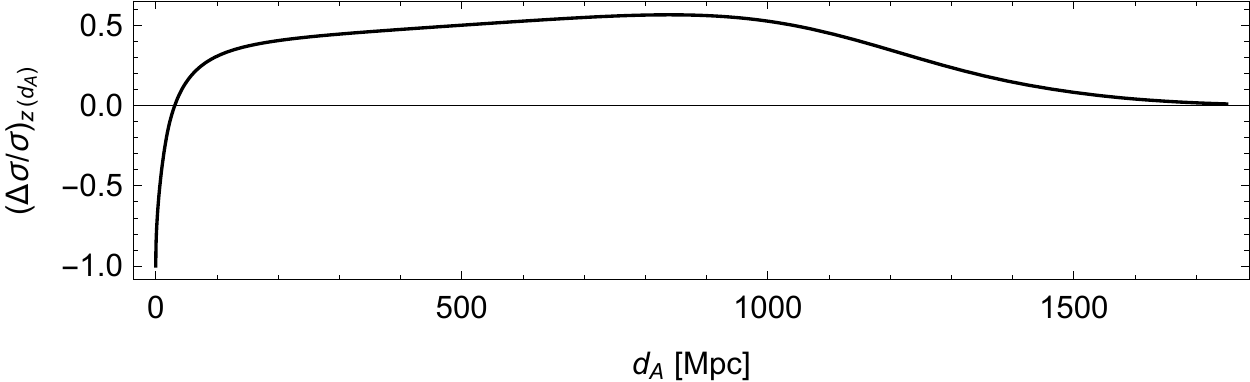}
\includegraphics[scale=0.6]{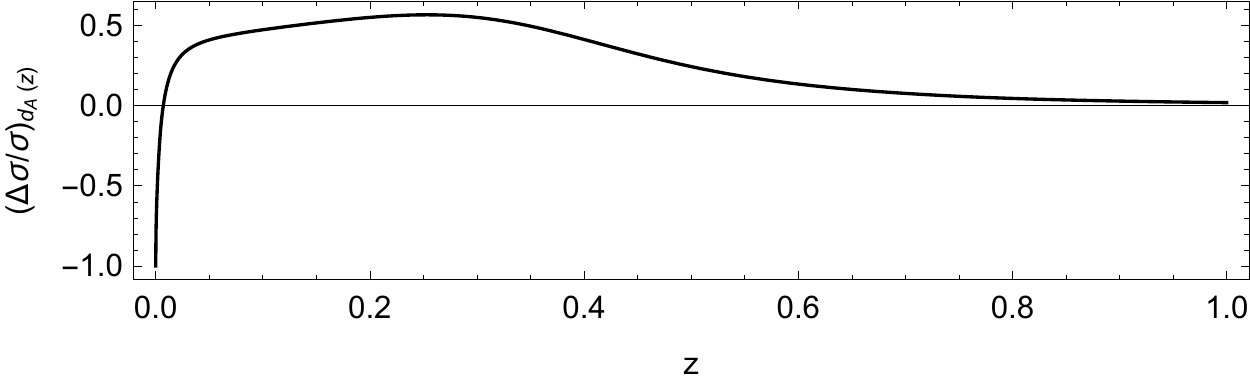}
\caption{Contribution to the dispersion from the observer velocity $v_{\rVert\,o}$ generated by cosmological perturbations along the past world-line of the observer. In the upper panels, there are the dispersion without including the terms from the $v_{\rVert\,o}$ (solid gray lines), the pure correction due to $v_{\rVert\,o}$ (dashed gray lines) and the sum of them (solid black lines). Lower panels show the relative correction to $\sigma$ from the observer motion. Left panels are for $z(d_A)$ while right ones refer to $d_A(z)$. It is evident that this effect is competitive with the lensing convergence and source Doppler one and it can cause a shift between $40\%$ and $60\%$ for younger/closer sources.}
\label{fig:obs}
\end{figure}

 The same properties appear also in the variances, where we get
\beq
\left(\sigma^2_{d_{A\,z}}\right)_{v_{\rVert\,o}}=\left(\frac{1}{\hat r_z\Hcal_z}\right)^2\overline{\langle v^2_{\rVert\,o} \rangle}\qquad,\qquad
\left(\sigma^2_{z_d}\right)_{v_{\rVert\,o}}=\left(\frac{1}{\hat r_d\Hcal_d-1}\right)^2\overline{\langle v^2_{\rVert\,o} \rangle}
\eeq
More in details, in Figs. \ref{fig:obs} we show the contribution to the dispersion due to the autocorrelation of the observer velocity and the cross correlation between observer and source velocities. In particular, bottom panels show the relative change in the total dispersion due to the effect of the observer peculiar velocity. It is impressive to notice that the total effect can be affected by a shift between 40 and 60 $\%$. This result in in agreement with \cite{Biern:2016kys}. However, for younger/closer sources, this contribution generates an almost perfect cancelation, leading to a $\sim -100\%$ change. This suppression is due to the cross-term of $\overline{\langle v_{\rVert} v_{\rVert\,o} \rangle}$ in the estimation of the variance. Indeed, this term can be written as (see Appendix \ref{app:tec})
\beq
\overline{\langle v_\rVert\,v_{\rVert\,o} \rangle}=\left( 1+z \right)^{-1}\,\mathcal{F}(\eta_s)\,\overline{\langle v^2_\rVert \rangle}
\label{eq:crossDoppler}
\eeq
with
\beq
\mathcal{F}(\eta_s)=\frac{\int \frac{dk}{k}\,k^2\,\Pcal_\Psi(k,\eta_o)j_0(k\left( \eta_o-\eta_s \right))}{\int \frac{dk}{k}\,k^2\,\Pcal_\Psi(k,\eta_o)}
\label{eq:F}
\eeq
and the integrals are made over the $k$-modes in Fourier space, $\Pcal_\Psi$ is the dimensionless power spectrum for the Bardeen potential and $j_0$ is the spherical Bessel function of $0$-th order. For high redshift, the spherical Bessel function modulates the integrand of the numerator in such a way that the ratio between the two integrals becomes significantly lower than one. Moreover, the whole term is further decreased by the inverse of the resdhift. On the other hand, for close/young sources when $\eta_s\rightarrow\eta_o$, $j_0\rightarrow 1$ and $1+z$ becomes negligible. In this way, the ratio between the two integrals is almost unity and then it follows that $\overline{\langle v_\rVert\,v_{\rVert\,o} \rangle}\approx\overline{\langle v^2_\rVert \rangle}$. This effect then is competitive with the auto-correlations of velocities at really small redshifts. In fact, this competitive behavior between $\overline{\langle v_\rVert\,v_{\rVert\,o} \rangle}$ and $\overline{\langle v^2_\rVert \rangle}$ exhibits a huge cancelation between these effects. These properties are shared between the $d_A(z)$ and $z(d_A)$ relations. More in details, for the angular-distance/redshift relation, the variance due to the cross-correlation of the velocities is
\beq
\left(\sigma^2_{d_{A\,z}}\right)_{v\,v_o}=2\,\frac{\hat r_z\Hcal_z-1}{(\hat r_z\Hcal_z)^2}\overline{\langle v_\rVert\,v_{\rVert\,o} \rangle}\,.
\label{eq:crossDopplersigma}
\eeq
Combining Eqs. \eqref{eq:crossDoppler} and \eqref{eq:crossDopplersigma}, we can write
\beq
\left(\sigma^2_{d_{A\,z}}\right)_{v\,v_o}=\frac{2}{1+z}\,\frac{\mathcal{F}(\hat \eta_z)}{\hat r_z\Hcal_z-1}\,\left(\sigma^2_{d_{A\,z}}\right)_{v_\rVert}\,.
\label{eq:vvoVSvv}
\eeq
For very small redshift, we can write
\begin{align}
\left(\sigma^2_{d_{A\,z\rightarrow 0}}\right)_{v_\rVert}^{\text{tot}}=&
\overline{\langle v^2_{\rVert\,o} \rangle}\left[ \frac{1}{(\hat r_z\Hcal_z)^2}+\left(\frac{\hat r_z\,\Hcal_z-1}{\hat r_z\,\Hcal_z}\right)^2
+2\,\frac{\mathcal{F}(\hat \eta_z)}{1+z}\frac{\hat r_z\,\Hcal_z-1}{\left(\hat r_z\,\Hcal_z\right)^2} \right]\nonumber\\
=&\overline{\langle v^2_{\rVert\,o} \rangle}\left[ 1
-2\,\left(1-\frac{\mathcal{F}(\hat \eta_z)}{1+z}\right)\frac{\hat r_z\,\Hcal_z-1}{\left(\hat r_z\,\Hcal_z\right)^2} \right]\,.
\end{align}
Looking at Eq. \eqref{eq:F}, for young sources, $j_0(k \hat r_z)\approx1-k^2 \hat r_z^2/6$, so we can write $\mathcal{F}(\hat \eta_z)=1-\mathcal{F}_2\,\hat r_z^2$, where $\mathcal{F}_2\sim 10^{-4}$ Mpc$^{-2}$ is a redshift independent constant. This means that
\beq
\left(\sigma^2_{d_{A\,z\rightarrow 0}}\right)_{v_\rVert}^{\text{tot}}=\overline{\langle v^2_{\rVert\,o} \rangle}\left( 1+2\,\frac{\mathcal{F}_2}{1+z}\frac{\hat r_z\,\Hcal_z-1}{\Hcal_z^2} \right)\approx \overline{\langle v^2_{\rVert\,o} \rangle}=(2.4\times 10^{-3})^2\,.
\label{eq:sigmaDAZobs}
\eeq
Hence, for very close sources, the total contribution due to the peculiar velocity of observer and source does not diverge as just $\left(\sigma^2_{d_{A\,z}}\right)_{v_{\rVert}}$ would do, but it remains finite and it is exactly $\overline{\langle v^2_{\rVert\,o} \rangle}$. The finite smallness of this term explains the contribution of $\sim-100\%$ in the right bottom panel of Fig. \ref{fig:obs}, where we get a finite result instead of a divergent one. This behavior is also in agreement with the limit of the $\delta d_{A\,z}$ for the source approaching the observer position. Indeed, looking at Eq. \eqref{eq:da}, for $s\rightarrow o$, we get that $\delta d_{A\,z}=v_{\rVert\,o}+\text{potentials}$, then its variance is in agreement with Eq. \eqref{eq:sigmaDAZobs} and takes into account the huge cancelation. This behavior is characteristic just for very close objects. Its consequence is to decrease the total amplitude of the variance close to the observer, as also numerically shown in Fig. \ref{fig:obs}. However, in the regime $0.01\le z\le 0.03$, it is not helpful in getting $\sigma^2_{d_{A\,z}}$ competitive with $\left(\sigma^{H_0}_{d_{A,z}}\right)^2$.
\begin{figure}[ht!]
\centering
\includegraphics[scale=0.6]{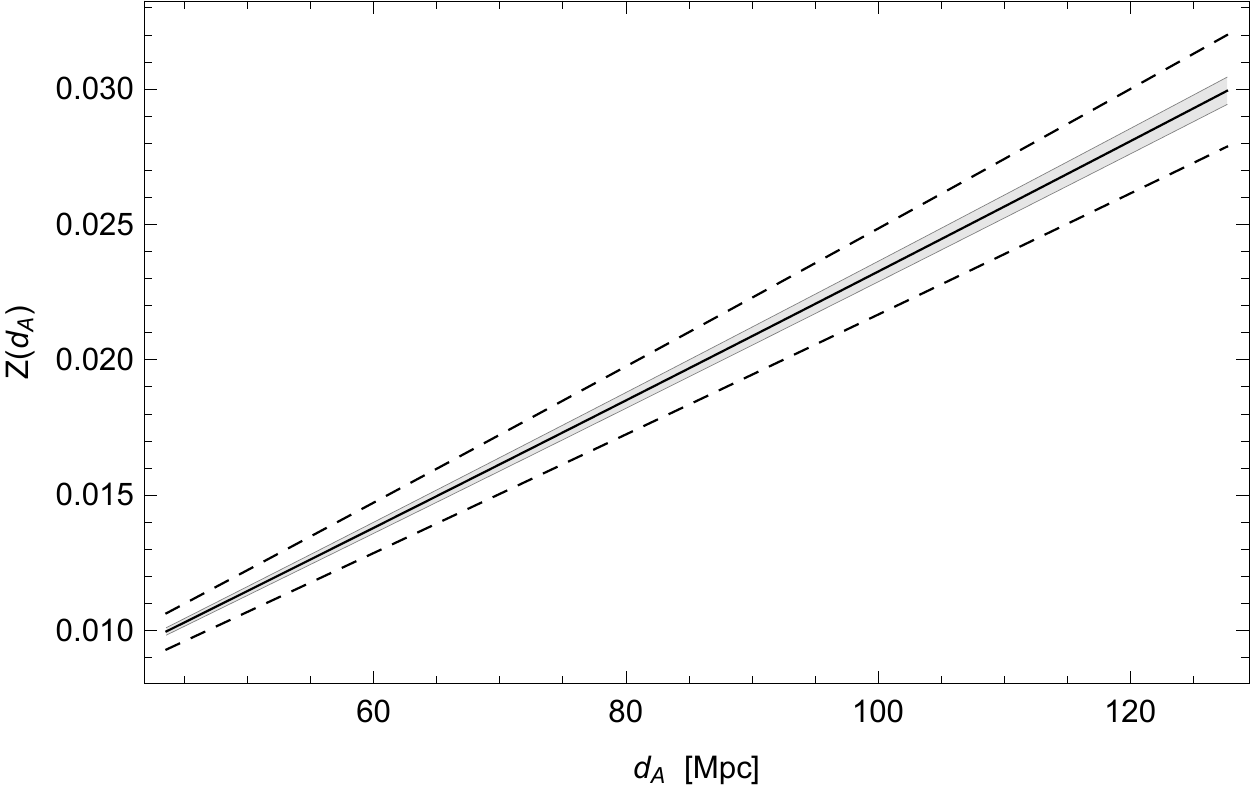}
\includegraphics[scale=0.6]{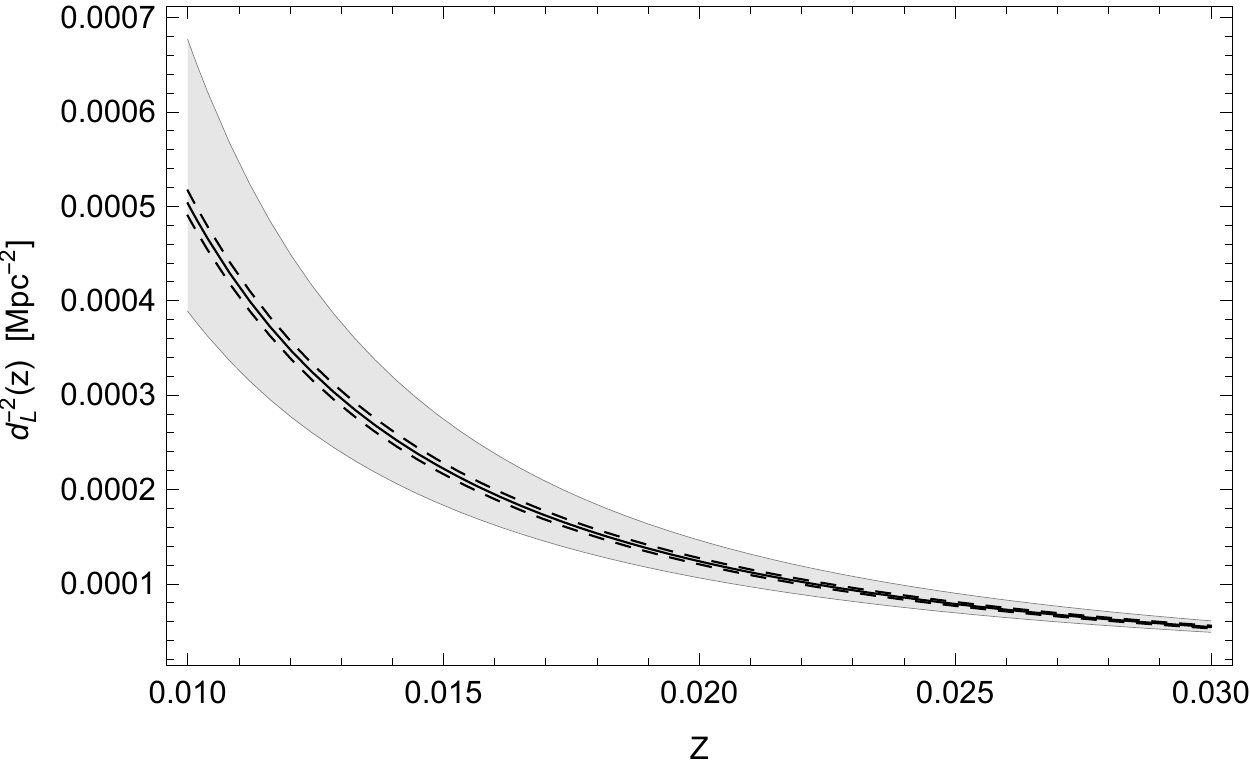}\\
\includegraphics[scale=0.6]{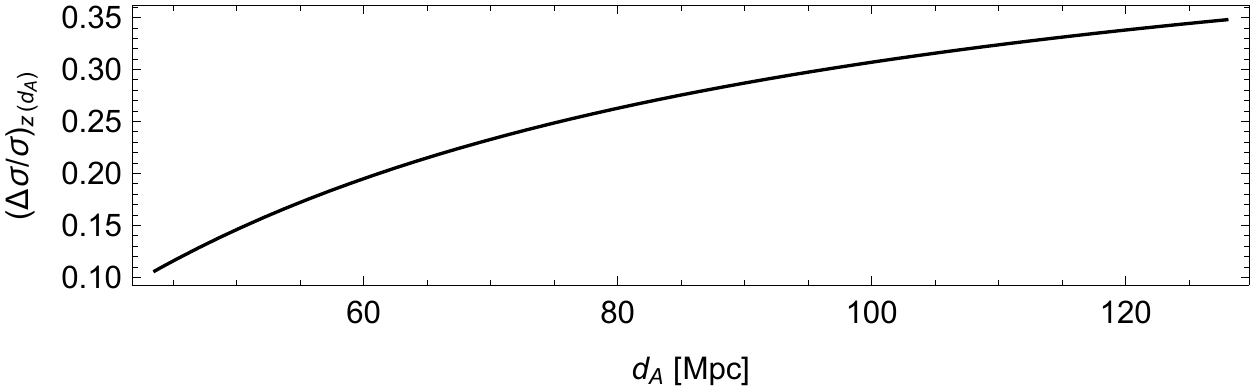}
\includegraphics[scale=0.6]{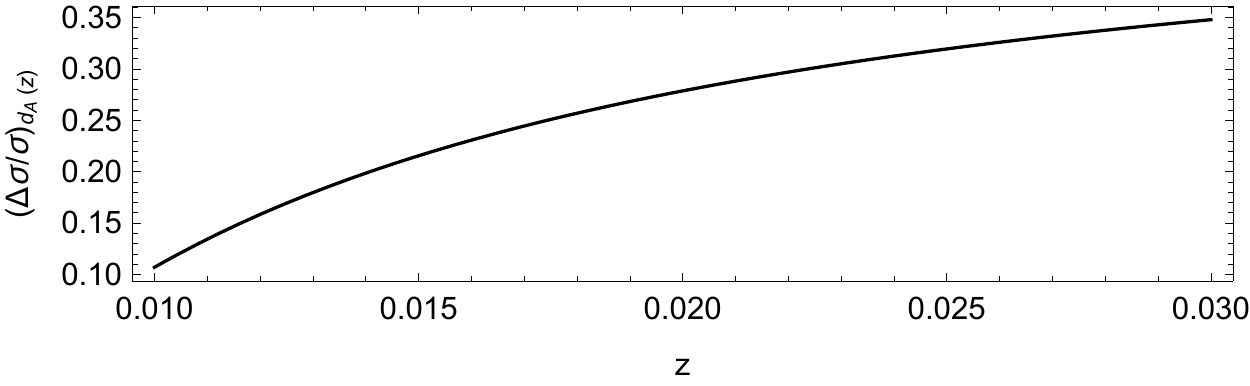}
\caption{Plots for the redshift $z$ in function of the observed angular-distance (top left panel) and observed flux $\Phi\equiv d_L^{-2}$ as function of the redshift (top right panel). Black solid curves are the expected background relations, with $H_0=67.8$ Km/s Mpc$^{-1}$, dashed lines are the same background curves where $H_0\pm\Delta H_0=67.8\pm 0.9$ Km/s Mpc$^{-1}$ and the gray regions are due to the theoretical dispersion caused by inhomogeneities where the contributions from $\kappa$, $v_\rVert$ and $v_{\rVert\,o}$ are taken into account. Note that the gray region and dashed lines in the left panel are for 10-$\sigma$ dispersion rather than 1-$\sigma$, in order to make them visible. Bottom panels show the relative change due to the observer velocity contribution in the dispersion for $z(d_A)$ (left) and for $d_L^{-2}(z)$.}
\label{fig:H0local}
\end{figure}

Indeed the top right panel in Figs. \ref{fig:H0local} compares the total dispersion of the flux to the inhomogeneities (gray region) with the dispersion allowed by the actual uncertainty of $H_0$ measured by Planck. As we can see in the bottom right panel, where the relative change in $\sigma$ due to $v_{\rVert\,o}$ is presented in the redshift range investigated in \cite{Ben-Dayan:2014swa}, the cancelation due to $v_{\rVert\,o}$ is still not so strong. On the contrary, there is an increase of almost $30\%$ of the total dispersion. Because of this, the analysis made in \cite{Ben-Dayan:2014swa} about the tension in the measurement of $H_0$-local and $H_0$-CMB not only remains valid but could also get a significant increase of the total theoretical systematic which affects the measure of $H_0$ from local supernovae, as already claimed in \cite{Ben-Dayan:2014swa}.

On the contrary, for the redshift/angular-distance the cancelation is exact. We can easily understand this by looking at the velocity terms in Eq. \eqref{eq:z}. In the limit where $s\rightarrow o$, $v_{\rVert\,s}$ and $v_{\rVert\,o}$ cancel so no contribution is expected from them. Indeed, with the same argument provided for $\delta d_{A\,z}$, we have that
\beq
\left(\sigma^2_{z_{d\rightarrow 0}}\right)_{v_\rVert}^{\text{tot}}=\overline{\langle v^2_{\rVert\,o} \rangle}\left[ \frac{1}{(\hat r_d\Hcal_d-1)^2}+1
+2\,\frac{\mathcal{F}(\hat \eta_d)}{1+\bar z_d}\frac{1}{\hat r_d\,\Hcal_d-1} \right]\,,
\label{eq:sigmaZDAobs}
\eeq
which manifestly goes to 0, when $\hat r_d\rightarrow 0$. More generally, we notice that the full expression for $\delta z_d$ goes to 0 when $s\rightarrow o$. This means that no contribution to the dispersion today is expected to appear.

\subsection{Potentials}
The last set of perturbations that may contribute to the dispersion of the two relations considered here is due to the local and integrated potentials. While their amplitudes are expected to be small, the major concern about their presence is about their infrared convergence in the integration over $k$ in Fourier space. Indeed, these terms do not involve any spatial derivative of the gravitational potential and this implies that they always contribute to the integral over $k$-modes in Fourier space as
\beq
\int\frac{dk}{k}\,\Pcal_\Psi(k)\,j_0(k\Delta \eta)=\int_0^{k_{IR}}\frac{dk}{k}\,\Pcal_\Psi(k)\,j_0(k\Delta \eta)+\int_{k_{IR}}\frac{dk}{k}\,\Pcal_\Psi(k)\,j_0(k\Delta \eta)\,,
\label{eq:truncated}
\eeq
where $\Delta\eta$ may be null, $\eta_o-\eta_s$ or considered as the difference of times along the integrated lines of sight. The choice of the infrared cut-off $k_{IR}$ is usually made such that it corresponds to the horizon scale, i.e. $\Hcal_0$. This choice has a physical meaning, because we do not want that super-horizon scales ($k\ll \Hcal_0$) affect the sub-horizon physics. Despite the fact that this cut-off is put by hand, we notice that in the super-horizon regime $\Pcal_\Psi(k)\sim k^{n_s-1}$. Moreover, the spherical Bessel function can be Taylor expanded as
\beq
j_0(x)=\sum_{n=0}^\infty(-1)^n\frac{x^{2n}}{(2n+1)!}
\eeq
so the truncated infrared part of the $k$ integral, namely the irst integral in the rhs of Eq. \eqref{eq:truncated}, can be written as
\beq
\sum_{n=0}^\infty\frac{(-1)^n}{(2n+1)!}\,\Delta\eta^{2n}\int_0^{k_{IR}}\,dk\,k^{2(n-1)+ns}
=\sum_{n=0}^\infty\frac{(-1)^n}{(2n+1)!}\,\Delta\eta^{2n}\left[ \frac{k^{2n+n_s-1}}{2n+n_s-1} \right]^{k_{IR}}_0\,.
\label{eq:series}
\eeq
For the red spectrum from Planck data \cite{Ade:2013zuv}, $n_s=0.96$. This means that the squared brackets in Eq. \eqref{eq:series} are finite for $n>0$ whereas a divergent behavior appears for $n=0$. Then, the final result exhibits a dependency on the infrared cut-off. However, it has been shown in \cite{Biern:2016kys} that the divergent behavior is canceled in the total estimation of the variance for the angular-distance/redshift relation thanks to the presence of $\delta \eta_o$. Indeed, the divergent part of Eq. \eqref{eq:series} ($n=0$) never depends on $\Delta\eta$. This implies that time and $k$ integrals factorize in the expression of the variance. Then, the entire contribution to $\sigma^2$ due to these divergent terms can be written as
\begin{align}
\left(\sigma^2_{d_{A\,z}}\right)_\text{IR}=&\left[ -\left( 2-\frac{1}{\hat r_z\Hcal_z} \right)g(\hat\eta_z)
+\frac{2}{\hat r_z}\int_{\hat\eta_z}^{\hat\eta_o}d\eta\,g
-\frac{1-\hat r_z\,\Hcal_z}{\hat r_z\,\Hcal_z}\,g(\hat\eta_o)
+2\,\frac{1-\hat r_z\,\Hcal_z}{\hat r_z\,\Hcal_z}\,
\int_{\hat\eta_z}^{\hat\eta_o}d\eta\,g'\right.\nonumber\\
&\left.-\frac{1}{a_o}\left(\Hcal_o-\frac{\Hcal_o}{\hat r_z\,\Hcal_z}+\frac{1}{\hat r_z}\right)\int_0^{\hat\eta_o}d\eta\, a\,g \right]^2\left[ \frac{k^{n_s-1}}{n_s-1} \right]^{k_{IR}}_0\,,
\label{eq:cornerstone}
\end{align}
where $g(\eta)$ is reported in Eq. \eqref{eq:growth}. It is the growth function of the gravitational potential such that $\psi(\eta)/g(\eta)=$ constant. In order to get a convergent result, we then need to prove that all the terms in the squared brackets sum up exactly to 0. Here we provide an analytical proof in the simpler case of the CDM solution, where $a=\left(\eta/\eta_o\right)^2$, $g(\eta)=1$ and $\Hcal\equiv a'(\eta)/a(\eta)=2/\eta$. With these conditions, we get that Eq. \eqref{eq:cornerstone} is
\begin{align}
\left(\sigma^2_{d_{A\,z}}\right)_\text{IR}=&\left[ -1
+\frac{2}{\hat r_z}\int_{\hat\eta_z}^{\hat\eta_o}d\eta\,
-\frac{1}{a_o}\left(\Hcal_o-\frac{\Hcal_o}{\hat r_z\,\Hcal_z}+\frac{1}{\hat r_z}\right)\int_0^{\hat\eta_o}d\eta\,\left( \frac{\eta}{\hat\eta_o} \right)^2 \right]^2\left[ \frac{k^{n_s-1}}{n_s-1} \right]^{k_{IR}}_0\nonumber\\
=&\left[ 1
-\frac{3}{\hat\eta_o}\,\int_0^{\hat\eta_o}d\eta\,\left( \frac{\eta}{\hat\eta_o} \right)^2 \right]^2\left[ \frac{k^{n_s-1}}{n_s-1} \right]^{k_{IR}}_0=0\,.
\end{align}
The integrals in these lines are exactly the contribution from $\delta\eta_o$. Without them, the variance would diverge as $k^{-0.04}$. This convergent result can be proven also for a purely cosmological constant dominated era in the expansion of the Universe. More in general, it happens whenever the pressure content in the energy-momentum tensor is zero or constant. In this regard an analytical proof of the cancellation of Eq. \eqref{eq:cornerstone} during the matter-cosmological constant phase transition has been provided in \cite{Biern:2016kys}.

The convergence of the variance for the redshift/angular-distance relation follows directly from the finiteness of $d_A(z)$. Indeed, by noticing that the variance for $z(d_A)$ and $d_A(z)$ are related by
\beq
\sigma^2_{z_d}=\left(\frac{\hat r_d\,\Hcal_d}{\hat r_d\,\Hcal_d-1}\right)^2\,\sigma^2_{d_{A\,z}}(\hat \eta_z=\hat \eta_d,\hat r_z=\hat r_d)\,,
\label{eq:relation}
\eeq
we get that the variance for $z_d$ shares the same infrared behavior as the one for $d_A(z)$. Indeed the overall factor in Eq. \eqref{eq:relation} goes to 0 when $\hat r_d\ll \Hcal_d^{-1}$ and $\sigma^2_{d_{A\,z}}$ remains finite. This means that the total variance for the redshift/angular-distance relation goes to zero for very close sources. Once the convergence of the whole expression has been provided, we then have that the contribution from the potential to the total dispersion in both relations is very negligible.

\section{Summary}
\label{sec:summary}
In this work we have investigated the relation between redshift and angular-distance at linear level in metric perturbations. This procedure depends on which kind of relation we want to consider, in order to correctly define our reference background. Indeed, for $d_A(z)$, our sources are lying on the constant redshift spheres. On the contrary, $z(d_A)$ considers sources which are located on constant angular-distance spheres. Because these two spheres are different, this means that these relations are not simply related by an analytical inverse function but they also have to consider displacements of the source position with respect to different observed reference quantities.

Having this in mind, in Sect. \ref{sec:linear} we have then provided the analytical derivation of the displacements with respect to the constant redshift and the constant angular-distance spheres as long as both of them are on the observed past light-cone. After that we have respectively applied them to the angular-distance/redshift relation and to the redshift/angular-distance one. It turns out that both relations involve the same relativistic effects as expected. More in details, the leading effect for older/fainter sources is the weak lensing whereas for young/close objects the source radial velocity is the relevant term. However, these relations do not share the same properties. Indeed for $z(d_A)$ the source velocity is not amplified by a divergent factor but remains almost constant for all distances.

This behavior exhibits very different properties also in the cosmological estimators. Indeed, in Sect. \ref{sec:variance} we have estimated the dispersions associated to $d_A(z)$ and $z(d_A)$ due to the presence of linear inhomogeneities in the Universe in the ranges of redshift and angular-distance of young/close sources. The interesting results that we get is that the dispersion associated to redshift/angular-distance relations stays constant for close sources and is $\sim 0.1\%$. This value is almost 2 orders of magnitude lower than the case of $d_A(z)$. Moreover its value is competitive with the actual error of $H_0$ from Planck. This is not the case for the angular-distance/redshift relation, where the dispersion is so large that it could explain the tension between the measurement of $H_0$-local and $H_0$-CMB. In this sense, the estimation of $H_0$ from the $z(d_A)$ relation rather than $d_A(z)$ should exhibits a theoretical systematics (or cosmic variance) small enough to provide a better comparison between this kind of $H_0$-local measurement and $H_0$-CMB.

The conclusions made so far remain valid also when other contributions from observer peculiar velocity and (local or integrated) gravitational potential are taken into account. Indeed, in Sect. \ref{sec:others} we shown that the observer peculiar motion due to cosmological perturbations can enlarge by $\sim 50\%$ the total dispersion for both relations in some intermediate range of redshift or angular-distance. Hence this contribution could amplify the cosmic variance associated to the $d_A(z)$ relation but is not large enough to avoid an unbiased estimation of $H_0$-local from $z(d_A)$. On the contrary, for very young/close sources the observer velocity terms may be important because they provide a cancellation of the total dispersion which is exact for the redshift/angular-distance relation. Still in Sect. \ref{sec:others} we have also commented about the fact that corrections due to the gravitational potential do not exhibit any divergent behavior. Then they remain small enough to be safely neglected in the analysis.

\section*{Acknowledgments}
I am very grateful to Giovanni Marozzi for highlighting some crucial points about this project. I also thank Fulvio Scaccabarozzi for helpful preliminary comparisons of numerical results. Moreover, I thank Maurizio Gasperini, Ermis Mitsou and Jaiyul Yoo for their comments about the final version of the draft. Finally I acknowledge support by the Swiss National Science Foundation and by a Consolidator Grant of the European Research Council (ERC-2015-CoG grant 680886).

\appendix
\section{Perturbative expressions}
\label{app:perturbed}
In this appendix, we report the expression for $\delta z$, $\delta w$ and $\delta d_A$ in terms of linear scalar metric perturbations of the longitudinal gauge
\beq
\delta g_{\mu\nu}=-2\,a(\eta)^2\left[\phi\,d\eta^2+\psi\,\left( dr^2+r^2\,d\Omega^2 \right)\right]\,.
\eeq
Moreover, we impose vanishing anisotropic stress, namely $\phi=\psi$. Then, our expressions are
\begin{align}
\delta z
=&\,\left[-\psi
-v_\rVert\right]^s_o
-2\int_{\eta_s}^{\eta_o}d\eta\,\psi'+\Hcal_o\,\delta\eta_o\qquad,\qquad
\delta \eta_o=-\frac{1}{a_o}\int_{\eta_{in}}^{\eta_o}d\eta\,a\,\psi\nonumber\\
\delta w=&-2\int_{\eta_s}^{\eta_o}d\eta\,\psi-\delta \eta_o-\delta r_o\qquad,\qquad\delta d_A=-v_{\rVert\,o}-\psi_s-\kappa-\frac{\delta r_o}{r}\,,
\end{align}
where $\delta r_o$ is a normalization function which drops out in the combination $\delta d_A-\delta w/r$.
Notice that the integrals between $\eta_s$ and $\eta_o$ are performed along the past light-cone, at $r=\eta_o-\eta$, while the integrals between $\eta_{in}$ and $\eta_o$ (or $\eta$) are performed along the world-line of the observer (or source), i.e. $r=0$ (or $r=\eta_o-\eta_s$). As already stated in the main text of this work, let us notice that $\delta z\sim v_{\rVert\,s}$ and $\delta d_A\sim\kappa$. These terms are explicitly given by
\beq
v_{\rVert}=\int_{\eta_{in}}^{\eta_s} d\eta\,\frac{a(\eta)}{a(\eta_s)}\pa_r\psi\quad,\quad\kappa=\frac{1}{\eta_o-\eta_s}\int_{\eta_s}^{\eta_o}d\eta\frac{\eta-\eta_s}{\eta_o-\eta}\Delta_2\psi\quad\text{and}\quad v_{\rVert\,o}=v_\rVert(\eta_s=\eta_o)
\eeq
where $\Delta_2$ is the angular Laplacian. Moreover, the total expressions for the desired relations are
\begin{align}
\delta d_{A\,z}=&-\frac{1}{\hat r_z\Hcal_z}v_{\rVert\,o}
-\left( 2-\frac{1}{\hat r_z\Hcal_z} \right)\psi_s
-\kappa
+\frac{2}{\hat r_z}\int_{\eta_s}^{\eta_o}d\eta\,\psi
+\frac{1-\hat r_z\,\Hcal_z}{\hat r_z\,\Hcal_z}\,\left(
v_{\rVert\,s}
-\psi_o\right)\nonumber\\
&+2\,\frac{1-\hat r_z\,\Hcal_z}{\hat r_z\,\Hcal_z}\,
\int_{\eta_s}^{\eta_o}d\eta\,\psi'
-\frac{1}{a_o}\left(\Hcal_o-\frac{\Hcal_o}{\hat r_z\,\Hcal_z}+\frac{1}{\hat r_z}\right)\,\int_{\eta_{in}}^{\eta_o}d\eta\,a\,\psi
\label{eq:da}
\end{align}
and
\begin{align}
\delta z_d=&\frac{\hat r_d\,\Hcal_d}{1-\hat r_d\,\Hcal_d}\,\kappa
-\frac{2\,\Hcal_d}{1-\hat r_d\,\Hcal_d}\,\int_{\eta_s}^{\eta_o}d\eta\,\psi
-\frac{1-2\,\hat r_d\,\Hcal_d}{1-\hat r_d\,\Hcal_d}\,\psi_s
+\psi_o\nonumber\\
&-v_{\rVert\,s}
+\frac{1}{1-\hat r_d\,\Hcal_d}\,v_{\rVert\,o}
-2\int_{\eta_s}^{\eta_o}d\eta\,\psi'
-\frac{1}{a_o}\left(\Hcal_o-\frac{\Hcal_d}{1-\hat r_d\,\Hcal_d}\right)\int_{\eta_{in}}^{\eta_o}d\eta\,a\,\psi\,.
\label{eq:z}
\end{align}
Eq. \eqref{eq:da} is in full agreement with the one presented in \cite{Sasaki:1987ad,Biern:2016kys,Scaccabarozzi:2017ncm}. Moreover, aside the last term due to the observer time lapse \cite{Biern:2016kys,Fanizza:2018qux}, we notice agreement for all the other terms even with \cite{Bonvin:2005ps,Fanizza:2013doa,Umeh:2012pn}.

\section{Gauge invariance of the relations between observables}
\label{app:GI}
Here we verify that both relations $d_A(z)$ and $z(d_A)$ are gauge-invariant, just as expected for relations between physical observables. Indeed, under a linear gauge transformation $\tilde x^\mu=x^\mu+\epsilon^\mu$ of the background sets of coordinates, we have that (see \cite{Fanizza:2018qux} for explicit derivation of these transformation properties)
\begin{align}
\widetilde{\delta z}=\delta z+\Hcal\,\epsilon^\eta-\Hcal_o\epsilon^\eta_o\qquad,\qquad
\widetilde{\delta \eta_o}=\delta \eta_o+\epsilon^\eta_o\qquad,\qquad
\widetilde{\delta w}=\delta w-\epsilon^\eta-\epsilon^r\,.
\label{eq:GT}
\end{align}
The gauge transformation for $\delta z$ follows from the fact that $z$ is a bi-scalar, so it gauge-transforms at both observer and source position. The gauge transformation for $\delta\eta_o$ is obtained from the fact that it is the expansion of the time-coordinate around the time as measured by the observer in its rest-frame $\hat\eta_o$, i.e. $\eta_o=\hat\eta_o+\delta\eta_o$. Then, because $\tilde\eta_o=\eta_o+\epsilon^\eta_o$ and $\hat\eta_o$ is required to be gauge-invariant, the transformation for $\delta\eta_o$ naturally follows. The gauge transformation for $\delta w$ is due to the fact that it is the perturbation of the past light-cone given by $\bar w=\eta+r$ on the background so, from its gauge transformation $\widetilde{\delta w}=\delta w-\epsilon^\mu\pa_\mu\bar w$, we get the last of Eqs. \eqref{eq:GT}. What remains to be evaluated is $\widetilde{\delta d_A}$. Again, because its background value is $\bar d_A=a(\eta)\,r$, we easily get
\beq
\widetilde{\delta d_A}=\delta d_A-\frac{\epsilon^\mu\pa_\mu\bar d_A}{\bar d_A}
=\delta d_A
-\Hcal\,\epsilon^\eta
-\frac{\epsilon^r}{r}\,.
\eeq
In this way, it is straightforward to show that both combinations in Eqs. \eqref{eq:DLZ} and \eqref{eq:DLDA} transforms as
\beq
\widetilde{\delta z+\Hcal_o\delta\eta_o}=\delta z+\Hcal_o\delta\eta_o+\Hcal\epsilon^\eta\qquad,\qquad
\widetilde{\delta d_A-\frac{\delta w}{\hat r}}=\delta d_A-\frac{\delta w}{\hat r}
-\frac{\hat r\,\Hcal-1}{\hat r\,\Hcal}\,\Hcal\epsilon^\eta\,,
\eeq
and then verify that the gauge field $\epsilon^\eta$ cancels in both $\delta d_{A\,z}$ and $\delta z_d$.

\section{Expressions in Fourier space and technical details}
\label{app:tec}
First of all, we express the $\psi$ in Fourier space as
\beq
\psi(\eta,\vec x)=\frac{1}{\left( 2\pi \right)^{3/2}}\int d^3 k\,E(\vec k)\,\psi_k(\eta)e^{i\,r\vec k\cdot\hat x}
\eeq
where $E(\vec k)$ is a random field such that $E^*(\vec k)=E(-\vec k)$. Moreover, we require that its statistical properties under the \textit{ensamble average} $\overline{\textcolor{white}{l}\dots\textcolor{white}{l}}$ satisfy $\overline{E(\vec k)}=0$ and $\overline{E(\vec k)E(\vec k')}=\delta\left( \vec k+\vec k' \right)$. In this way, we get that
\begin{align}
\pa_r\psi(\eta,\vec x)=&\frac{1}{\left( 2\pi \right)^{3/2}}\int d^3 k\,E(\vec k)\psi_k(\eta)\,i\,\vec k\cdot\hat x\,e^{i\,r\vec k\cdot\hat x}\nonumber\\
\Delta_2\psi\left( \eta,\vec x \right)=&-\frac{1}{\left(2\pi\right)^{3/2}}\int d^3k\,E(\vec k)\psi_k\left( \eta \right)\left( k^2r^2\,\sin^2\theta+2\,i\,kr\,\cos\theta \right)e^{i\,r\vec k\cdot\hat x}\,.
\end{align}
Hence the variance for the given quantity $\Ocal(\psi)$ can be computed as
\beq
\sigma^2_\Ocal\equiv\overline{\langle\Ocal(\psi)^2\rangle}=\overline{\frac{1}{4\pi}\int d^2\Omega\,\Ocal(\psi)^2}\,
\eeq
where $\langle\dots\rangle=\frac{1}{4\pi}\int d^2\Omega\dots$ is the \textit{average over directions}. Moreover, we consider the linear power spectrum of the Bardeen potential
\beq
|\psi_k(\eta)|^2=\frac{2\pi^2}{k^3}\,\Pcal_\Psi(k,\eta)
\eeq
where
\beq
\Pcal_\Psi(k,\eta)=A\,\left( \frac{3}{5} \right)^2\,\left( \frac{g(\eta)}{g_\infty} \right)^2\,\left( \frac{k}{k_0} \right)^{n_s-1} T^2\left(\frac{k}{13.41\,k_\text{eq}}\right)
\label{eq:power_spectrum}
\eeq
is the dimensionless power spectrum. Here $T(k)$ is the transfer function which takes into account the sub-horizon evolution of modes re-entering during the radiation domination era, here approximated by the Hu and Eisenstein \cite{Eisenstein:1997ik} parametrization
\begin{align}
T(q)=\frac{L_0(q)}{L_0(q)+q^2\,C_0(q)}\qquad&,\qquad
L_0(q)=\log(2\,e+1.8\,q)\nonumber\\
C_0(q)=14.2+\frac{731}{1+62.5\,q}\qquad&,\qquad k_\text{eq}=0.07\,h^2\,\Omega_{m0}
\label{eq:Hu}
\end{align}
and $g(\eta)$ is the growth function for the Bardeen potential satisfying the evolution equation
\beq
g''+3\,\Hcal\,g'+\left( 2\Hcal'+\Hcal^2 \right)\,g=0\,,
\eeq
here approximated by
\beq
g(\eta)=\frac{5}{2}\,g_\infty\frac{\Omega_m}{\Omega^{4/7}_m-\Omega_\Lambda+\left( 1+\frac{\Omega_m}{2} \right)\left( 1+\frac{\Omega_\Lambda}{70}\right)}\qquad\text{with}\qquad\Omega_m+\Omega_\Lambda=1
\label{eq:growth}
\eeq
where $g_\infty$ is a normalization constant chosen such that $g(\eta_o)=1$. The parameters in Eq. \eqref{eq:power_spectrum} and \eqref{eq:Hu} are given by \cite{Ade:2013zuv}
\beq
A=2.2\times10^{-9},\quad n_s=0.96,\quad k_0=0.05\,\text{Mpc}^{-1},\quad
h=0.678,\quad\Omega_{m0}=0.315\,.
\eeq
Hence, after some integration, we get our desired results
\begin{align}
\overline{\langle v^2_\rVert \rangle}=&\frac{1}{3}\,\left(\int_{\eta_{in}}^{\eta_s} d\eta\,\frac{a(\eta)g(\eta)}{a(\eta_s)g(\eta_o)}\right)^2\,\int \frac{dk}{k}\,k^2\,\Pcal_\Psi(k,\eta_o)\nonumber\\
\overline{\langle \kappa^2 \rangle}
=&\frac{1}{2}\,\int_{\eta_s}^{\eta_o}d\eta\int_{\eta_s}^{\eta_o}d\eta'
\frac{\left(\chi_s-\chi\right)\left(\chi_s-\chi'\right)}{\chi_s^2}
\frac{g(\eta)g(\eta')}{g^2(\eta_o)}
\int \frac{dk}{k}\,k^4\,\Pcal_\Psi(k,\eta_o)\nonumber\\
&\times\,\left[\chi\chi'\,\Ical_1\left(k\left( \eta'-\eta \right)\right)
+\frac{2}{k}\left( \eta'-\eta \right)\,\Ical_2\left(k\left( \eta'-\eta \right)\right)
-\frac{4}{k^2}\,\Ical_3\left(k\left( \eta'-\eta \right)\right)\right]\nonumber\\
\overline{\langle v_\rVert\,v_{\rVert\,o} \rangle}=&-\frac{3}{2}\,\frac{a(\eta_s)}{a(\eta_o)}\,\overline{\langle v^2_\rVert \rangle}\,\frac{\int \frac{dk}{k}\,k^2\,\Pcal_\Psi(k,\eta_o)\,\Ical_3(k\left( \eta_o-\eta_s \right))}{\int \frac{dk}{k}\,k^2\,\Pcal_\Psi(k,\eta_o)}\,.
\label{eq:sigmak}
\end{align}
where we have defined $\chi(\eta)\equiv\eta_o-\eta$ and $\Ical_n$ functions are enlisted here
\begin{align}
\Ical_1(x)=&48\,\frac{\sin x}{x^5}-48\,\frac{\cos x}{x^4}-16\,\frac{\sin x}{x^3}
\nonumber\\
\Ical_2(x)=&12\,\frac{\sin x}{x^4}-12\,\frac{\cos x}{x^3}-4\,\frac{\sin x}{x^2}
\nonumber\\
\Ical_3(x)=&4\,\frac{\sin x}{x^3}-4\,\frac{\cos x}{x^2}-2\,\frac{\sin x}{x}
\end{align}
It is interesting to notice that $\Ical_n$ can be exactly decomposed in terms of the spherical Bessel functions $j_n(x)\equiv(-x)^n\left( \frac{1}{x}\frac{d}{dx} \right)^n\frac{\sin x}{x}$. After a few algebraic manipulations, we find
\begin{align}
\Ical_1(x)=&\frac{16}{35}\,j_4(x)+\frac{32}{21}\,j_2(x)+\frac{16}{15}\,j_0(x)\nonumber\\
\Ical_2(x)=&\frac{3}{5}\, j_3(x)+\frac{3}{5}\,j_1(x)\nonumber\\
\Ical_3(x)=&\frac{4}{3}\,j_2(x)-\frac{2}{3}\,j_0(x)\,.
\end{align}
These relations allow us to approximate Eq. \eqref{eq:sigmak} by taking the highest number of powers in $k$. Indeed, by roughly counting $j_n(x)\sim \frac{\sin x}{x^{n+1}}$, the leading contribution can be addressed to $j_0$ in $\Ical_1$ and $\Ical_3$. Hence Eqs. \eqref{eq:sigmak} become
\begin{align}
\overline{\langle \kappa^2 \rangle}=&\frac{8}{15}\,\int_{\eta_s}^{\eta_o}d\eta\int_{\eta_s}^{\eta_o}d\eta'\frac{\left(\chi_s-\chi\right)\left(\chi_s-\chi'\right)\chi\chi'}{\chi_s^2}\frac{g(\eta)g(\eta')}{g^2(\eta_o)}
\int \frac{dk}{k}\,k^4\,\Pcal_\Psi(k,\eta_o)
\,j_0\left(k\left( \eta'-\eta \right)\right)\nonumber\\
\label{eq:approxkappa}
\end{align}
and
\beq
\overline{\langle v_\rVert\,v_{\rVert\,o} \rangle}=\left( 1+z \right)^{-1}\overline{\langle v^2_\rVert \rangle}\,\frac{\int \frac{dk}{k}\,k^2\,\Pcal_\Psi(k,\eta_o)j_0(k\left( \eta_o-\eta_s \right))}{\int \frac{dk}{k}\,k^2\,\Pcal_\Psi(k,\eta_o)}
\eeq
The analytical approximation done in Eq. \eqref{eq:approxkappa} looks in quite good agreement with the result presented in \cite{Fleury:2016fda}. First of all, they consider the power spectrum for density rather than the one for the Bardeen potential and this explains the different powers of $k$ in the integrals. Indeed the density $\delta$ and the gravitavional potential $\psi$ are related through the Poisson equation which, in Fourier space, reads as $k^2\psi_k\sim \delta_k$. This implies that $|\psi_k|^2\sim|\delta_k|^2\,k^{-4}$ and this provides the agreement between us and \cite{Fleury:2016fda} with respect to the power counting of $k$ in the integrals. However we notice that, in order to get Eq. \eqref{eq:approxkappa}, we have neglected several terms in the expressions of $\Ical_n$. This may lead to some small numerical discrepancies between our results and their ones.

\bibliographystyle{JHEP}
\bibliography{biblio}
\end{document}